\documentclass[twocolumn]{aastex631}
\usepackage{graphicx} % Allows including images
\usepackage{amsmath}
\usepackage{amsthm}
\usepackage{amssymb}
\usepackage{hyperref}
\usepackage{multirow}
\usepackage{booktabs}

\begin{document}

\title{Optimizing the Resolution of Hydrodynamic Simulations for MCRaT Radiative Transfer Calculations}

\author[0009-0001-9356-5400]{Jose Arita-Escalante}
\affiliation{Southeastern Universities Research Association, Washington, D.C. 20005, USA}
\affiliation{Astrophysics Science Division, NASA Goddard Space Flight Center,Greenbelt, MD 20771, USA}
\affiliation{Center for Research and Exploration in Space Science and Technology, NASA/GSFC, Greenbelt, Maryland 20771, USA}

\author[0000-0002-4299-2517]{Tyler Parsotan}
\affiliation{Center for Space Science and Technology, University of Maryland Baltimore County, 1000 Hilltop Circle, Baltimore, MD 21250, USA}
\affiliation{Astrophysics Science Division, NASA Goddard Space Flight Center,Greenbelt, MD 20771, USA}
\affiliation{Center for Research and Exploration in Space Science and Technology, NASA/GSFC, Greenbelt, Maryland 20771, USA}

\author[0000-0003-1673-970X]{S. Bradley Cenko}
\affiliation{Astrophysics Science Division, NASA Goddard Space Flight Center,Greenbelt, MD 20771, USA}

\affiliation{Joint Space-Science Institute, University of Maryland, College Park, MD 20742, USA}

%% Note that the \and command from previous versions of AASTeX is now
%% depreciated in this version as it is no longer necessary. AASTeX 
%% automatically takes care of all commas and "and"s between authors names.

%% AASTeX 6.31 has the new \collaboration and \nocollaboration commands to
%% provide the collaboration status of a group of authors. These commands 
%% can be used either before or after the list of corresponding authors. The
%% argument for \collaboration is the collaboration identifier. Authors are
%% encouraged to surround collaboration identifiers with ()s. The 
%% \nocollaboration command takes no argument and exists to indicate that
%% the nearby authors are not part of surrounding collaborations.

%% Mark off the abstract in the ``abstract'' environment. 

\begin{abstract}

Despite their discovery about half a century ago, the Gamma-ray burst (GRB) prompt emission mechanism is still not well understood. Theoretical modeling of the prompt emission has advanced considerably due to new computational tools and techniques. One such tool is the PLUTO hydrodynamics code, which is used to numerically simulate GRB outflows. PLUTO uses Adaptive Mesh Refinement to focus computational efforts on the portion of the grid that contains the simulated jet. Another tool is the Monte Carlo Radiation Transfer (MCRaT) code, which predicts electromagnetic signatures of GRBs by conducting photon scatterings within a jet using PLUTO. The effects of the underlying resolution of a PLUTO simulation with respect to MCRaT post-processing radiative transfer results have not yet been quantified. We analyze an analytic spherical outflow and a hydrodynamically simulated GRB jet with MCRaT at varying spatial and temporal resolutions and quantify how decreasing both resolutions affect the resulting mock observations. We find that changing the spatial resolution changes the hydrodynamic properties of the jet, which directly affect the MCRaT mock observable peak energies. We also find that decreasing the temporal resolution artificially decreases the high energy slope of the mock observed spectrum, which increases both the spectral peak energy and the luminosity. We show that the effects are additive when both spatial and temporal resolutions are modified. Our results allow us to understand how decreased hydrodynamic temporal and spatial resolutions affect the results of post-processing radiative transfer calculations, allowing for the optimization of hydrodynamic simulations for radiative transfer codes.

\end{abstract}

%% Keywords should appear after the \end{abstract} command. 
%;'//% The AAS Journals now uses Unified Astronomy Thesaurus concepts:
%% https://astrothesaurus.org
%% You will be asked to selected these concepts during the submission process
%% but this old "keyword" functionality is maintained in case authors want
%% to include these concepts in their preprints.


%% From the front matter, we move on to the body of the paper.
%% Sections are demarcated by \section and \subsection, respectively.
%% Observe the use of the LaTeX \label
%% command after the \subsection to give a symbolic KEY to the
%% subsection for cross-referencing in a \ref command.
%% You can use LaTeX's \ref and \label commands to keep track of
%% cross-references to sections, equations, tables, and figures.
%% That way, if you change the order of any elements, LaTeX will
%% automatically renumber them.
%%
%% We recommend that authors also use the natbib \citep
%% and \citet commands to identify citations.  The citations are
%% tied to the reference list via symbolic KEYs. The KEY corresponds
%% to the KEY in the \bibitem in the reference list below. 

\section{Introduction} \label{sec:intro}

A number of different theories have been created to understand the phenomena of Gamma-ray bursts (GRBs) since their initial discovery in the 1960's \citep{1973klebesadel}. One of the earliest models used to explain GRB prompt emission was the Synchroton Shock Model (SSM)\citep{1994meszaros}, which considers radiation generated when shells with different Lorentz factors collide with each other outside of the photospheric region \citep{2011daigne}. The collisions of these shells create perturbations of the magnetic fields that lead to the excitation of leptons which then emit synchroton radiation. The SSM can naturally explain GRB properties such as lightcurve variability and the observed nonthermal spectra. Nevertheless, it fails to agree with observed correlations of GRBs such as the Amati and Yonetoku relations \citep{2002amati, 2004yonetoku, 2011zhang-yan}.

Another model explaining the prompt emission mechanism is the photospheric model, which explains the phenomenon by describing thermal radiation that originates deep within a relativistic jet \citep{2005meszaros}. The radiation is initially in a part of the jet with a high optical depth, leading to many interactions between the photons and the matter in the jet. As the jet expands, it becomes optically thin, allowing photons to leave the jet's photosphere and travel to the observer without additional interactions with the GRB jet. The photospheric model is able to reproduce correlations that the SSM cannot, but is unable to replicate non-thermal spectral low and high-energy tails without the consideration of the photospheric region \citep{2010beleborodov, 2008peer, 2011peer-ryde} and subphotospheric dissipation events \citep{2015chhotray-lazzati}.

With the aid of computational tools, we have been better able to understand the physics of GRBs. Previous studies have conducted rigorous radiative transfer calculations, however; they have assumed that the jet structure has been simplified into an analytic profile \citep{2013ito,2014ito,2016vurm}. In contrast to radiative transfer calculations, other studies have utilized hydrodynamic (HD) calculations to simulate complex jet structures, but these only provide information about the matter within the jet \citep{2009lazzati, 2013lazzatiA, 2014lopez-camara}, which leads to a lack of information regarding the evolution of the radiation. 

The state of the art method to account for both of these assumptions is to perform post-processing radiative transfer calculations on a hydrodynamic (HD) simulated jet using Monte Carlo methods. There have been tools developed to perform post-processing radiative transfer calculations such as the ones developed by \cite{2015ito,2019ito} and the Monte Carlo Radiation Transfer (MCRaT) code \citep{2016lazzati,2018parsotanA,2018parsotanB,parsotan2021photosphericA,parsotan2021photosphericB}. MCRaT was developed to conduct radiative transfer calculations on HD simulations to generate mock observations of simulated GRBs using the photospheric model. The impact that HD simulation resolutions have on MCRaT post-processing radiative transfer calculations has not been studied yet. Ensuring that radiative transfer calculations are converged and accurate is critical to testing GRB prompt emission theories against observations.

Here, we present an analysis of HD resolution and its effect on post-processing radiative transfer calculations for simulated GRB mock observables. Section \ref{sec:methods} outlines the code used to create the HD jet, the code used to perform the radiative transfer calculations, and the way in which resolutions are quantified. Sections \ref{sec:results} and \ref{sec:discussion} show the effect that HD simulation resolutions have on radiative transfer calculations and the physical implications of these results. 

\section{Methods} \label{sec:methods}

In this section, we outline the methods to our analysis. In Section \ref{methods_sec:codes}, we discuss the codes used in our analysis. In Section \ref{sec:convergence}, we quantify convergences in our simulations.

\subsection{Codes Used}\label{methods_sec:codes}

Here, we discuss the codes used in our study. Section \ref{methods_subsec:pluto} highlights the tools used to create the numerical HD GRB simulation. In Section \ref{methods_subsec:mcrat}, we discuss the tools used to conduct post-processing radiative transfer calculations and analyze the results to generate mock observables for simulated GRBs.

\subsubsection{PLUTO} \label{methods_subsec:pluto}

%\begin{itemize}
%    \item PLUTO HD numerical solver
%    \item PLUTO AMR (can be very brief, should focus on the simulation, based on a 16TI progenitor star, etc)
%    \item CHOMBO simulation HD properties convergence (include plot) (information on this sim is 16TI in a Lazzati et.al.)
%\end{itemize}

\begin{figure*}[t!]
    \includegraphics[width = \linewidth]{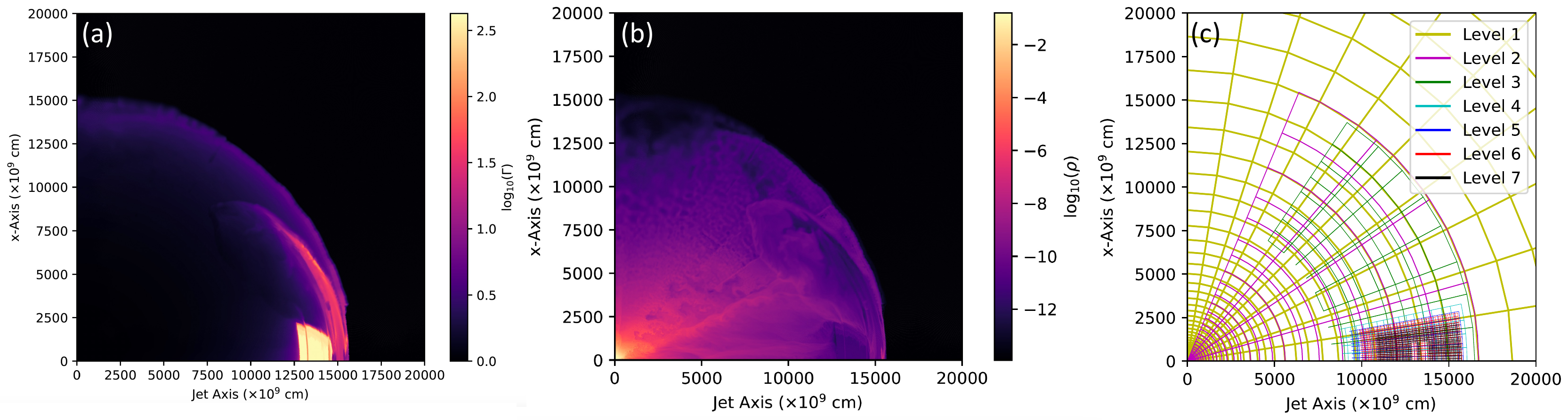}
    \caption{Frame properties at the end of the PLUTO 16TI simulation. These quantities are read at time given by $t_{sim} = \frac{\text{frame}}{\text{fps}}=527.6$ s. Panel (a) shows the Lorentz factor profile at this time. Panel (b) shows the mass density profile at this time. Panel (c) shows the AMR Grid shape and levels at this time. In this last panel, we can see the computational efforts are dedicated to the higher refinement area of the frame, where the jet is located. We note that even though Panel (c) shows 7 levels of refinement, in our analysis we highlight the 5 highest levels of spatial refinement in any HD frame.}
    \label{fig:hdplots}
\end{figure*}

PLUTO is a numerical solver for systems of partial differential equations in the context of astrophysical fluid dynamics \citep{2007mignone}. PLUTO uses CHOMBO Adaptive Mesh Refinement (AMR) to focus computational efforts on the most relevant parts of the HD simulation \citep{2012mignone}. Here, we used the PLUTO hydrodynamics code with AMR to simulate the propagation of a long gamma-ray burst (LGRB) jet from a 16TI stellar progenitor, taken from \cite{woosley2006progenitor}. The stellar progenitor profile was interpolated onto the PLUTO grid using the code's capability to do such operations. Following the prescription provided by \cite{lazzati2013photospheric}, we inject the jet with a constant luminosity of  $5.33 \times 10^{50}$ erg $\text{s}^{-1}$ for 100 s from an
injection radius of $1 \times 10^9$ cm, with an initial Lorentz factor of
5, an opening angle $\theta_0 = 10^\circ$, and an internal over rest-mass
energy ratio, $\eta = 80$. The simulation is given in a 2D spherical coordinate system. The simulation domain in PLUTO is logarithmic in radius from $1 \times 10^9$ cm to $5.6 \times 10^{14}$ cm, with 1600 cells in the radial direction, although the simulation is only carried out until the jet head reaches $\sim 2 \times 10^{13}$ cm. It also extends in polar angle from $0^\circ$ to $90^\circ$ covered within 160 cells. The AMR refinement is set such that the jet is followed with a resolution of at least $1 \times 10^9$ cm along the jet axis, leading to a maximum of 7 refinement levels needed for the AMR. The state of the jet is saved with a frame rate of 5 frames per second. 

A snapshot of the simulation can be found in Figure \ref{fig:hdplots}. These plots show a snapshot taken at the end of the simulation, at $t_{sim} = 527.6$ s. Figure \ref{fig:hdplots}(a) shows the Lorentz factor as a function of position in the frame. Figure \ref{fig:hdplots}(b) shows the mass density profile of this last frame. Figure \ref{fig:hdplots}(c) shows the AMR grid for this frame where level 1 is the base hydrodynamic grid that was previously defined. The highest level of refinement is level 7 in this figure which is focused on the core of the jet.

\begin{figure}
    \centering
    \includegraphics[width = \linewidth]{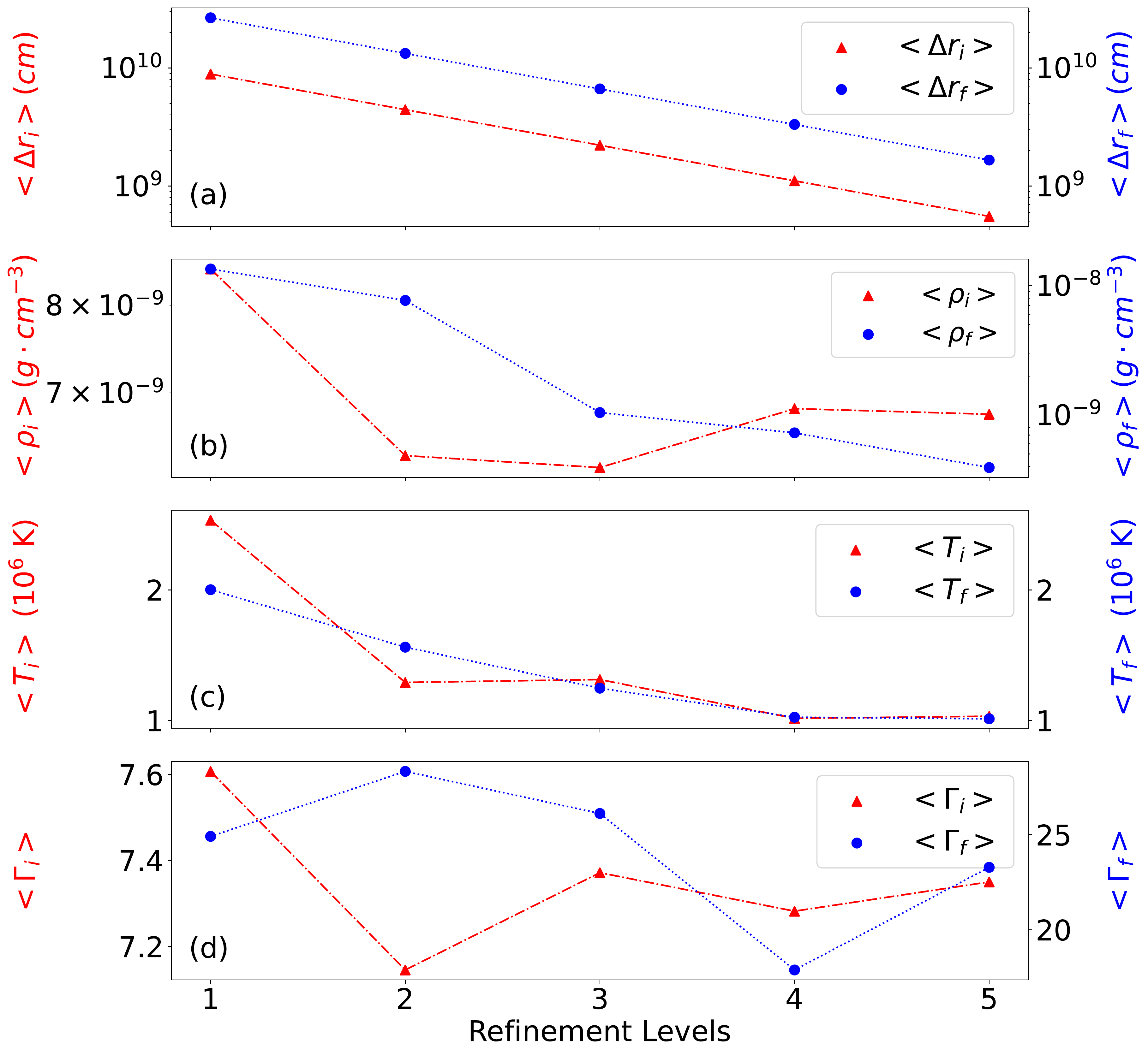}
    \caption{HD properties at different refinement levels for the PLUTO 16TI simulation as functions of refinement level at the first moment of photon injection and final frame of the simulation. We took the initial frame measurements of a shell located in radius $r = 1.3\times 10^{12}$ cm and time $t_{sim} = \frac{\text{frame}}{\text{fps}} = 50$ s. We took the final frame measurements of a shell located in radius $r = 1.5\times 10^{13}$ cm and time $t_{sim} = \frac{\text{frame}}{\text{fps}} = 527.6$ s. Panel (a) shows the average HD cell radius size ($\Delta r$) in cm. Panel (b) shows the average density in $\text{g cm}^{-3}$. Panel (c) shows the average temperature in K. Panel (d) shows the average bulk Lorentz factor. The early-time conditions are shown as red triangle markers and the final conditions are shown in blue circle markers. Throughout our analysis we focus on the 5 highest spatial refinement levels in a given HD simulation frame. Thus, the final frame quantities for refinement level 5 shown in this figure correspond to the black grid representing level 7 shown in \ref{fig:hdplots}(c). }
\label{fig:hdprops}
\end{figure}

We show the refinement and the convergence of the hydrodynamic properties of the simulation at the initial moment of photon injection and at the end of the simulation in Figure \ref{fig:hdprops}. 

We select a shell of grid cells around a radius of $1.3 \times 10^{12}$ cm at 50 s in the simulation and show the spatial resolution that is achieved at each refinement level at that time. We also show the convergence of the bulk Lorentz factor, $\Gamma$, density, $\rho$, and temperature , $T$\footnote{Temperature is calculated assuming that the jet is radiation dominated.}, as we traverse the different refinement levels. This convergent behavior is present throughout the whole simulation. In Figure \ref{fig:hdprops} we also show the HD properties of a shell of grid cells located at $1.5 \times 10^{13}$ cm at 527.6 s in the simulation, which is the last frame of our simulation. As is outlined in Section \ref{methods_subsec:mcrat} we focus on the 5 highest spatial refinement levels in the simulation regardless of the hydrodynamic simulation frame that is read in. Thus, the final frame quantities for refinement level 5 shown in Figure \ref{fig:hdprops} correspond to the black grid representing level 7 shown in \ref{fig:hdplots}(c).

\subsubsection{MCRaT and ProcessMCRaT}\label{methods_subsec:mcrat}

The MCRaT\footnote{\href{https://github.com/lazzati-astro/MCRaT}{https://github.com/lazzati-astro/MCRaT}} code conducts radiative transfer calculations to compute the electromagnetic (EM) signature of HD simulated GRB jets. 
MCRaT reads in HD simulations of GRB jets and performs Compton scatterings between the injected photons and matter in the jet. The code first reads in the simulation snapshots and injects photons within a specified region of the jet, where the number of injected photons depends on the comoving temperature of the jet. These photons are then propagated between and within each subsequent snapshot of the HD simulation \footnote{There are no discretized emission regions for photons in the MCRaT calculations as there are in other systems where monte carlo radiative transfer methods are used (e.g. stars in \cite{sph_rt} or \cite{parsotan_galaxy_rt}). There are mechanisms that allow for continuous photon emission such as that from cyclo-synchrotron emission \citep{parsotan2021photosphericA} however we do not consider that mechanism in this study. Additionally, subphotospheric photon mediated shocks are expected to produce photons in GRB jets but these physics are not yet captured in large scale, global HD or radiative transfer simulations \citep{ito_rms}}. The post-processing radiative transfer method that MCRaT employs is necessary for conducting photon transport calculations in special relativity within a medium that is time dependent.

MCRaT can run two different radiative transfer calculations. The first one is based off of reading in an HD numerical simulation of a GRB jet. In our study, we use the PLUTO 16TI simulation mentioned in Section \ref{methods_subsec:pluto}. The outflow given by this numerical simulation introduces numerically induced errors in the HD properties of the grid, so the effect of HD resolution on the post-processing radiative transfer calculations can be assessed. The other type of radiative transfer calculation MCRaT is capable of running is one of an analytic spherical outflow, or fireball. In the spherical outflow case, MCRaT takes the HD simulation files and overwrites the HD properties with those of an analytic outflow with only outward radial velocity components. This analytic outflow is a function of cell radius in which the spherical outflow is accelerated until an asymptotic Lorentz factor is reached. By using an analytic spherical outflow, we can understand how just the HD resolution has an effect on MCRaT mock observables. 

We set up our spherical outflow to have an asymptotic Lorentz factor $\Gamma_\infty = 100$, luminosity $L = 10^{54} \text{ erg}$ $\text{s}^{-1}$ and saturation radius $r_0 = 10^8 \text{ cm}$.

%\tyler{NOTE: We need to also add info about the angle range in which we injected photons, the totoal number of photons that we simulate, and the fact that we simulate photons within the first n seconds of the simulation. (in the 16TI case do we really only inject photons for 10 seconds? WE should do the full 100 seconds which is how long the jet is on for in the hydro simulation.  }

Spatial resolution for the HD simulation takes the form of various AMR refinement levels. Since the PLUTO AMR simulation that we conducted dynamically changes the number of refinement levels in order to maintain a resolution element size of $\sim 10^{9}$ cm, MCRaT reads in the $n$th highest refinement level at any given frame. Thus, level 5 is the highest refinement level at any time in the HD simulation, level 4 would be the second highest level, all the way until level 1, which is the lowest refinement level. We run multiple MCRaT simulations where we specify which of these spatial refinement levels should be read in from the PLUTO simulation. 

As outlined in Section \ref{methods_subsec:pluto}, the PLUTO simulation we use has a framerate of 5 frames per second (fps). For the context of our analysis, this is our highest temporal refinement level. We artificially vary the frame rate of our PLUTO simulation by telling MCRaT to only read every $n$th frame. With this method, we can achieve our desired framerate while maintaining the total simulation time for the HD simulation. In order to keep the same simulation time, each step in time ($\Delta t_{sim}$) has a specific number of frames assigned to it. As the resolution is lowered, the first and last frame in each $\Delta t_{sim}$ stay the same while the number of frames in between these two is lowered. This leads to a ``choppy" simulation.  In order to keep the simulation realistic as a function of time, each pair of subsequent frames varies more significantly as the temporal resolution is lowered. Reducing the framerate by a factor of 2 each time would be analogous to contiguous spatial refinement level HD cell radius increasing by a factor of 2. Therefore, 5 fps would be analogous to spatial refinement level 5, 2.5 fps would be analogous to spatial refinement level 4 and so on. This gives us a way to align the spatial refinement levels and come up with a clear way to mix and match temporal and spatial refinement levels, allowing us to investigate the effects of these changes combined with and independent of one another.

For all our MCRaT simulations, we kept our parameters as constant as possible. We injected photons into the HD simulation at an angle range of $0^\circ - 9^\circ$ and radius at $\sim 10^{12}$ cm.
We simulate photons within the first $100$ s of the PLUTO 16TI simulation, the time for which the GRB jet is active.
Additionally, we simulated $\sim 10^5 - 10^6$ photons per MCRaT simulation.

In order to analyze the output of MCRaT's simulations, we used ProcessMCRaT\footnote{\href{https://github.com/parsotat/ProcessMCRaT}{https://github.com/parsotat/ProcessMCRaT}} \citep{2021processmcrat}. ProcessMCRaT is a Python library developed to analyze and manipulate the output of MCRaT radiative transfer calculations. ProcessMCRaT fits the mock observed spectrum with a Band function \citep{1993Band} to calculate its low and high energy slopes, $\alpha$ and $\beta$ respectively, and its \replaced{peak energy, $E_{\text{pk}}$}{break energy, $E_{\text{0}}$. The peak  energy of the spectrum can then be calculated as $E_{\text{pk}}=(2+\alpha)E_{\text{0}}$ }. ProcessMCRaT also has the capability to create mock lightcurves for MCRaT simulated GRBs. ProcessMCRaT uses the photons in the last HD frame of the simulation to calculate mock observables, which corresponds to $t_{sim}= 527.6$ s for the PLUTO simulation used in this study which is shown in Figure \ref{fig:hdplots}. The photons are assumed to be in an optically thin regime of the jet and thus are no longer scattering with the matter in the jet, changing their energies, as they propagate to the observer. Choosing to construct mock observables at an earlier time in the simulation would lead to incorrect transient mock observables instead of steady state mock observables expected from photons that are no longer coupled to the jet. 

In producing our mock observables, we placed a mock observer at $r_{\text{obs}} = 10^{14}$ cm at various angles $\theta_{\text{obs}} = 1^{\circ}, 3^\circ, 5^\circ$ and $8^\circ$ from the GRB jet axis. The opening angle for the area in which the observer detects photons in $\Delta \theta_{\text{obs}} = 4^\circ$. We set the spectral fit for the observables to be that of a Band function including all photons at an energy range of $0.1$ - $4000$ keV.

We numerically integrated spectra with respect to energy to get luminosities, $L_\mathrm{iso}$. We also numerically integrated lightcurves with respect to time to obtain total isotropic energies, $E_\mathrm{iso}$.

%\begin{itemize}
%    \item Explain what ProcessMCRaT is, jupyter notebooks I created to reference github, etc
%    \item Mention the functions I used
%    \begin{itemize}
%        \item plot spectrum 
%        \item plot lightcurve
%        \item dictionary keys such as spectrum dict.e pk
%        \item instead of explicitly mentioning each function, I can just say I calculated spectra, lightcurves, etc using ProcessMCRaT.
%    \end{itemize}
%    \item ``we numerically integrated spectra" and how quantities worked out
%\end{itemize}

\subsection{Quantifying Convergence Within Radiative Transfer Calculations}\label{sec:convergence}

Since there are two dimensions of change in refinement (spatial and temporal), we populate a $5\times 5$ matrix that contains entries for its spatial and temporal refinement levels.

In order to quantify convergence in MCRaT mock observables between one spatial/temporal level and another, we define the percent change variable $\zeta_{\text{Prop}}$ as:
\begin{flalign}\label{eq:zeta_general}
        \nonumber\zeta_{\text{Prop}} (\text{lev}(n), & \text{fps}) = \\
        &\left|\frac{\text{Prop}(\text{lev}(n), \text{fps})-\text{Prop}(\text{lev}(5), 5\text{ fps})}{\text{Prop}(\text{lev}(5), 5\text{ fps})}\right|.    
\end{flalign}

Equation \ref{eq:zeta_general} represents a comparison of any level of temporal and spatial refinement with the highest combination of these both (spatial refinement 5 and 5 fps in our context) for any particular property (called Prop in Equation \ref{eq:zeta_general}) of the GRB EM signature, such as $\alpha$, $E_\mathrm{pk}$, or $L_\mathrm{iso}$\footnote{The calculations can be found here: \href{https://doi.org/10.5281/zenodo.8139769}{https://doi.org/10.5281/zenodo.8139769}}. 

This gives us a way to quantify the deviation at each level compared to the highest level for each GRB EM property. For the analysis of our results, the quantity $\zeta_{\text{Prop}}$ will be used to quantify deviations in resulting mock observables at different refinement levels.

\section{Results} \label{sec:results}

Here, we outline the results of our findings for the spherical outflow case and the 16TI simulation as described in Section \ref{sec:methods}. Our analysis shows the same trends for our mock observer angle $\theta_{\text{obs}}$ at all angles mentioned in Section \ref{methods_subsec:mcrat}. As a result, this section will only focus on $\theta_{\text{obs}}=1^\circ$.

\subsection{Spherical Outflow}\label{sec:spherical}

%\begin{itemize}
%    \item LC
%    \item spectra
%    \item e pk comparison plot (fps or spatial res)
%    \item L iso comparison plot (fps or spatial res)
%    \item E iso comparison plot (fps or spatial res)
%\end{itemize}

\subsubsection{Spectra}\label{results_subsec:spherical_spectra}

\begin{figure}
    \centering
    \includegraphics[width = \linewidth]{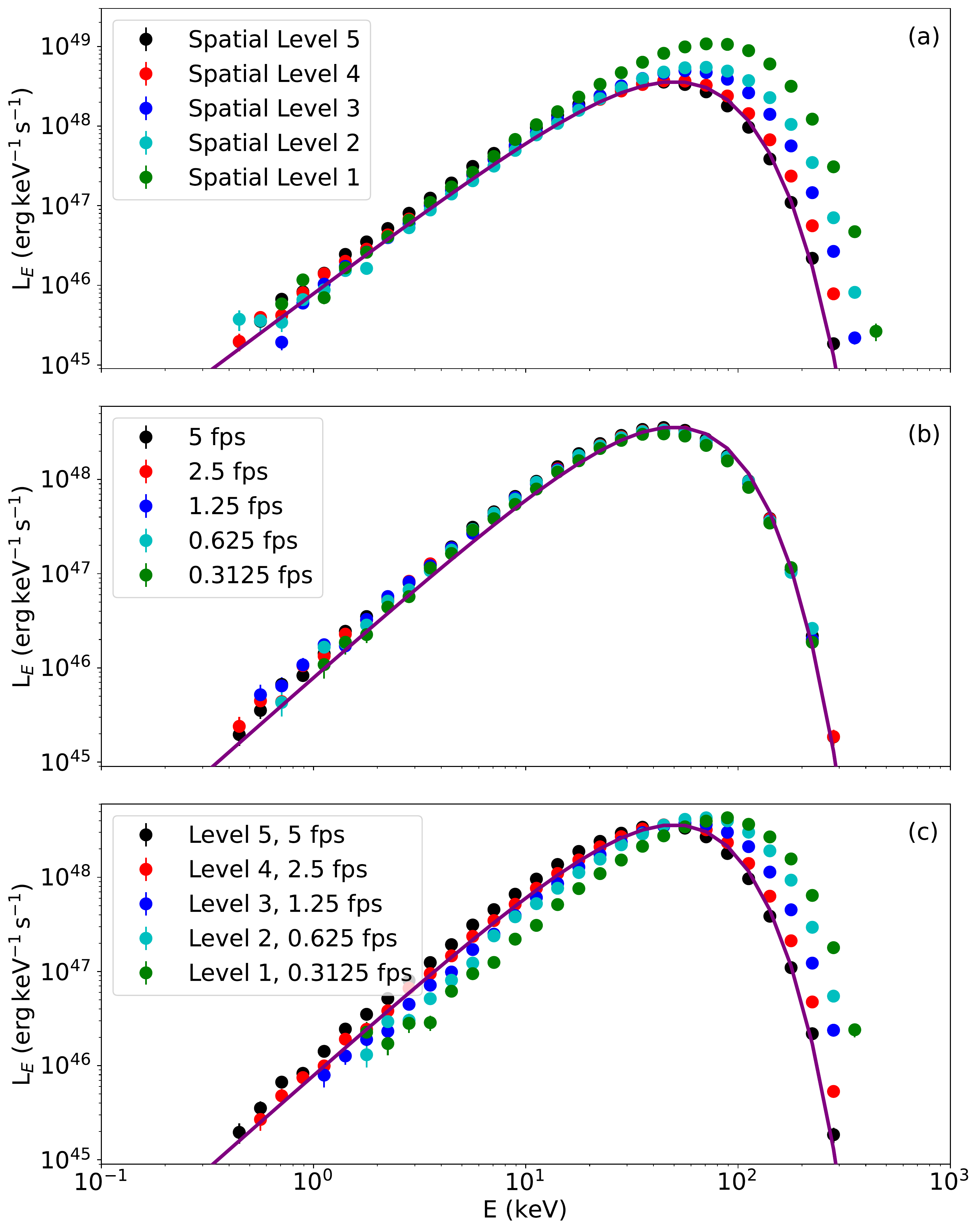}
    \caption{Spectra of a spherical outflow for different refinement levels. The solid purple line is a blackbody spectrum that peaks at each spectrum set's highest refinement level. Panel (a) shows the spectra of a spherical outflow profile, but at different spatial refinement levels while maintaining the same highest temporal resolution constant. Panel (b) shows the spectra generated with analytic outflow at different temporal resolutions, maintaining the highest spatial resolution constant. Panel (c) shows the spectra with matching temporal and spatial resolution levels.}
    \label{fig:spectra_spherical}
\end{figure}

Figure \ref{fig:spectra_spherical} shows spectra of a spherical outflow at different spatial and temporal resolutions. Figure \ref{fig:spectra_spherical}(a) shows spectra at the highest temporal resolution and varying spatial resolutions. Figure \ref{fig:spectra_spherical}(b) shows spectra at the highest spatial resolution and varying temporal resolutions. Figure \ref{fig:spectra_spherical}(c) shows spectra at matching temporal and spatial resolution levels.

\begin{figure}
    \centering
    \includegraphics[width = \linewidth]{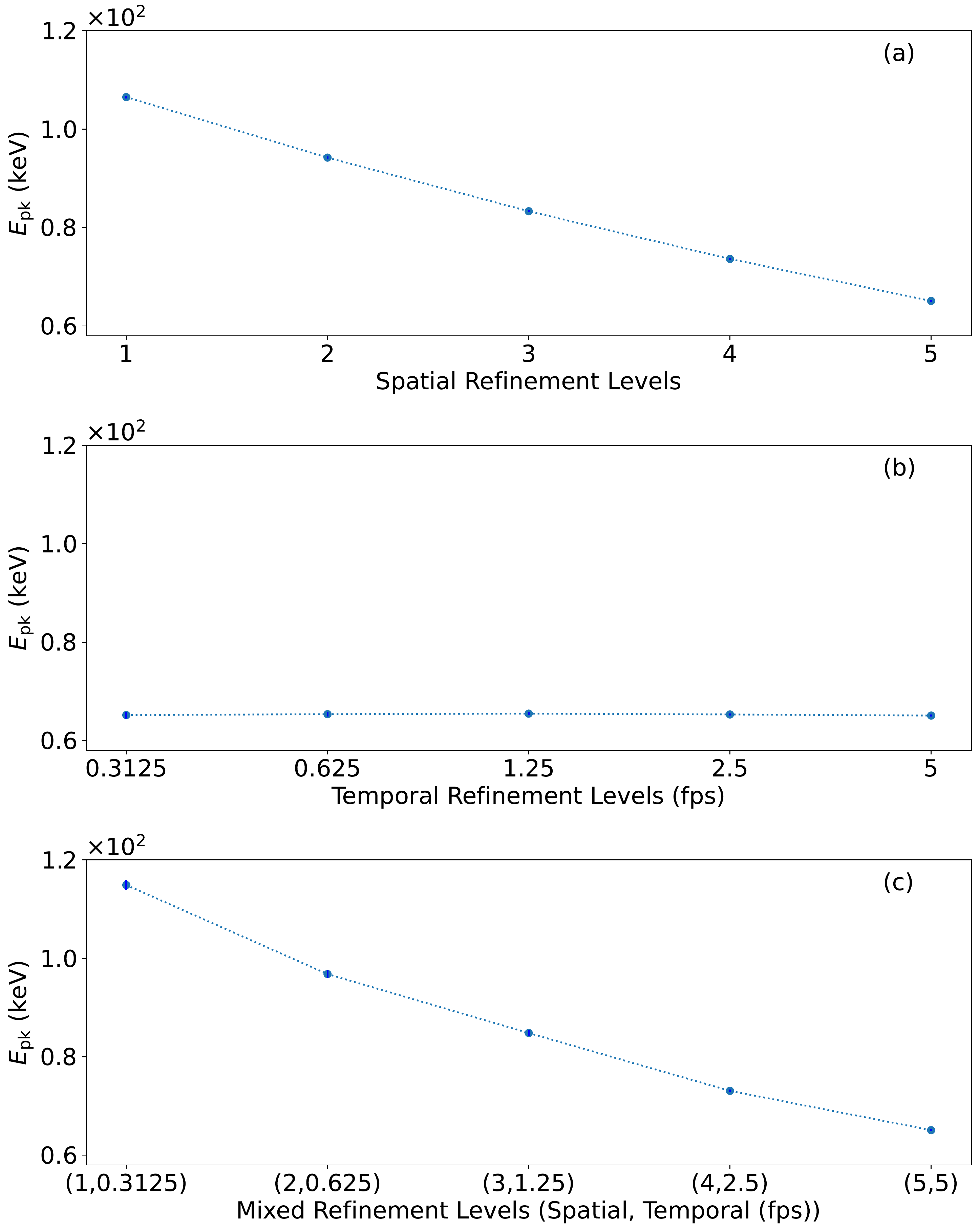}
    \caption{Spectral Peak energies of a spherical outflow for different refinement levels. Panel (a) shows the $E_{\mathrm{pk}}$ of the same spherical outflow simulation, but at different spatial refinement levels while maintaining the same highest temporal resolution constant. Panel (b) shows the $E_{\mathrm{pk}}$ generated with the same analytic outflow simulation at different temporal resolutions, maintaining the highest spatial resolution constant. Panel (c) shows the $E_{\mathrm{pk}}$ with matching temporal and spatial resolution levels. The error bars in panels (a), (b) and (c) are present, but are encompassed within the markers.}
    \label{fig:epk_spherical}
\end{figure}

When reducing the spatial resolution, we see an artificial increase of the peak energy of the spectrum. The higher HD cell sizes in lower resolutions cause the injection coordinates for photons to have different HD properties. As seen in Figure \ref{fig:hdprops}, lower spatial resolution levels have higher temperatures, which means the injected photons will have higher energies, causing a higher $E_{\mathrm{pk}}$. Figure \ref{fig:epk_spherical} shows the spectral peak energies at varying temporal and spatial resolutions. The effect of lowering the spatial resolution on spectral peak energies can be seen in Figure \ref{fig:epk_spherical}(a).

Reducing the spatial resolution causes an increase in the luminosity of the lightcurves. This happens since there is now more energy in the spectrum. The spectra then are shifted up in luminosity and to the right in energies, while still maintaining a blackbody shape. This effect can also be seen in Figure \ref{fig:epk_spherical}(a).

Reducing the temporal resolution of the simulation does not affect the spectrum of the spherical outflow in any significant manner. Since there is no change in spatial resolution, the injected photons read in the same HD values regardless of the temporal resolution. This effect can be seen in Figure \ref{fig:spectra_spherical}(b). For this reason, the peak energies are not affected as the temporal resolution is decreased, as seen in Figure \ref{fig:epk_spherical}(b). The luminosity of the spectra seems to be slightly decreased as the temporal resolution decreases. There is not a lot of variation in the spectral shape and properties since the homologous expansion present in a spherical outflow does not depend on time. 

Mixing spatial and temporal resolutions shows similar trends to only changing spatial resolutions. This is to be expected, since the analytic spherical outflow is defined to be time-independent. This effect can be seen in Figure \ref{fig:spectra_spherical}(c).

Analyzing the peak energies $E_{\mathrm{pk}}$ for different refinement levels confirms what we observed in the spectra in Figure \ref{fig:spectra_spherical}. As seen in Figure \ref{fig:epk_spherical}(a), there is an increase in peak energy as the spatial refinement level is decreased. This is seen as a shift to the right in the spectra in Figure \ref{fig:spectra_spherical}(a). For temporal resolutions, as seen in Figure \ref{fig:spectra_spherical}(b), there is no significant change between levels. This can be seen in the $E_{\mathrm{pk}}$ values in Figure \ref{fig:epk_spherical}(b) where the values are similar to one another. When mixing spatial and temporal resolutions, there is an additive behavior in the differences at various levels.

\begin{figure}
    \centering
    \includegraphics[width = \linewidth]{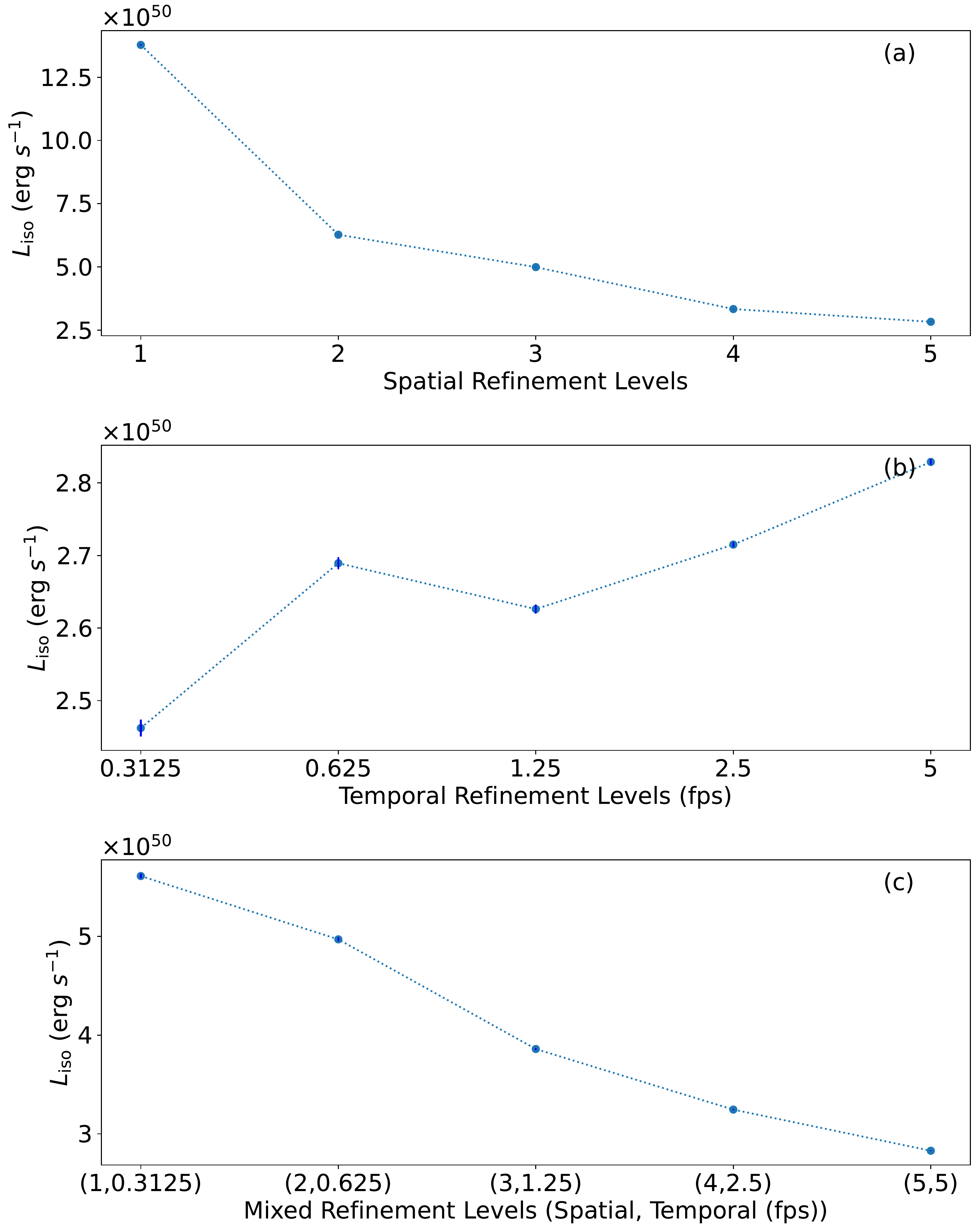}
    \caption{Spectral luminosities of a spherical outflow for different refinement levels. Panel (a) shows the luminosities using the same analytic outflow simulation, but at different spatial refinement levels while maintaining the same highest temporal resolution constant. Panel (b) shows the luminosities generated with the same spherical outflow simulation at different temporal resolutions while holding the highest spatial resolution constant. Panel (c) shows the luminosities with matching temporal and spatial resolution levels. The error bars in panels (a), (b) and (c) are present, but are encompassed within the markers.}
    \label{fig:lum_spherical}
\end{figure}

One other important quantity to observe is the luminosity at different spatial and temporal refinement levels. Figure \ref{fig:lum_spherical} shows the spectral luminosity at varying spatial and temporal resolutions with a spherical outflow. For differing spatial refinement levels, there is an artificial increase in luminosity as seen as a shift upwards in the spectra in Figure \ref{fig:spectra_spherical}(a). This effect can be seen in Figure \ref{fig:lum_spherical}(a). For differing temporal resolutions, luminosities tend to oscillate not too far way from each other as seen in Figure \ref{fig:lum_spherical}(b). This aligns with the very similar spectra seen in Figure \ref{fig:spectra_spherical}(b). When there is a mix of temporal and spatial refinement levels, luminosities also are artificially increased as the resolution goes down. This is the effect of the same phenomenon happening in spatial resolutions. The additive effect of combining spatial and temporal resolutions can be seen in Figure \ref{fig:lum_spherical}(c).

\subsubsection{Lightcurves}\label{subsec:spherical_lightcurves}

\begin{figure}
    \centering
    \includegraphics[width = \linewidth]{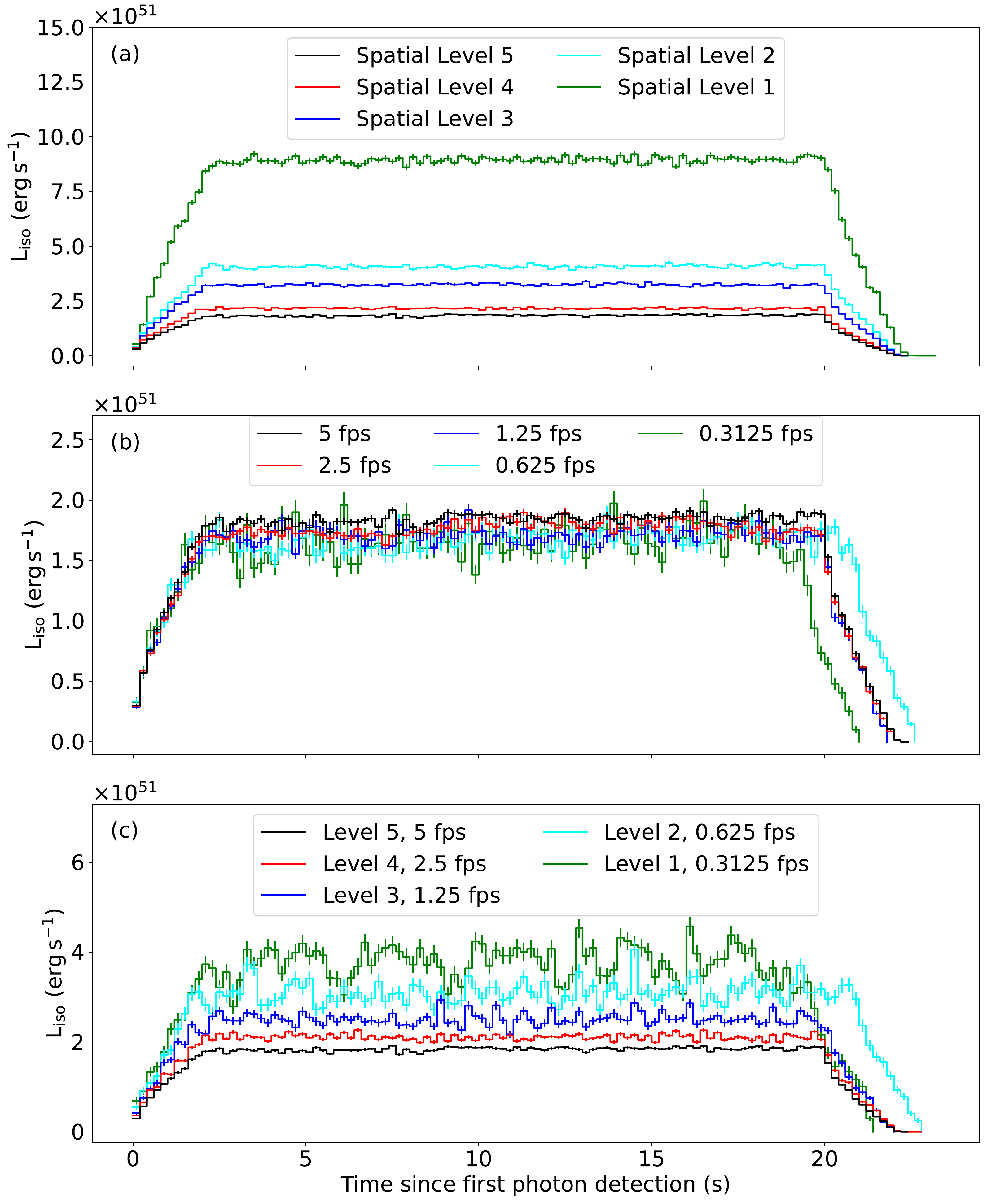}
    \caption{Lightcurves of simulated spherical outflow for different refinement levels. Panel (a) shows the lightcurves using the same analytic outflow simulation, but at different spatial refinement levels while maintaining the same highest temporal resolution constant. Panel (b) shows the lightcurves generated with the same spherical outflow simulation at different temporal resolutions, maintaining the highest spatial resolution constant. Panel (c) shows the lightcurves with matching temporal and spatial resolution levels.}
    \label{fig:lightcurves_spherical}
\end{figure}

Figure \ref{fig:lightcurves_spherical} shows lightcurves of a spherical outflow at different spatial and temporal resolutions. Figure \ref{fig:lightcurves_spherical}(a) shows lightcurves at the highest temporal resolution and varying spatial resolutions. Figure \ref{fig:lightcurves_spherical}(b) shows lightcurves at the highest spatial resolution and varying temporal resolutions. Figure \ref{fig:lightcurves_spherical}(c) shows lightcurves at matching temporal and spatial resolution levels.

In a spherical outflow, the lightcurves at different temporal and spatial resolutions have roughly the same shape. If the spatial resolution is decreased, as seen in Section \ref{results_subsec:spherical_spectra}, the luminosity of the spectrum is increased by an upwards shift of the spectrum. This causes lightcurves to have a higher luminosity. Analytic spherical outflows should have relatively ``constant" lightcurves. This can be seen in Figure \ref{fig:lightcurves_spherical}(a).

Reducing the temporal resolution does not change the luminosity of the lightcurves like spatial resolutions do. All temporal resolutions seem to roughly have the same luminosity. Analyzing the lightcurves, there is an increase in variability as the lightcurves oscillate around one ``average" value of the lightcurve. This variation is due to the fact that the photons are not being smoothly injected in a thin shell. Instead, the MCRaT algorithm has to determine which HD cells are the most energetic within a larger set of HD cells and correspondingly place more photons in those photon dense regions of the HD simulation. This leads to us only probing the portions of the outflow with the largest energies. This causes us to no longer get a smooth stream of photons that are detected as a function of time. This effect can be seen in Figure \ref{fig:lightcurves_spherical}(b).

Like with the spectra, varying both resolutions at the same time has an additive effect on the changed properties of the lightcurves. Figure \ref{fig:lightcurves_spherical}(c) shows how decreasing both the temporal and spatial resolution increases the luminosity of the lightcurve as well as the variability in the form of an oscillation around the ``average" value of the lightcurve.

\subsection{16TI HD Simulation}

%\begin{itemize}
%    \item LC \red{show figure with 3 panels, spatial, temporal and mixed res}
%    \item spectra \red{show figure with 2 panels, spatial and temporal}
%    \item e pk comparison plot (mixed)
%    \item L iso comparison plot (mixed)
%    \item E iso comparison plot (mixed)
%    \item spatial and temporal resolution matrix $\zeta$ plots [$\alpha$ , %$\beta$ , e pk, e iso, l iso ](all matrices in one page)
%\end{itemize}

\subsubsection{Spectra}

\begin{figure}
    \centering
    \includegraphics[width = \linewidth]{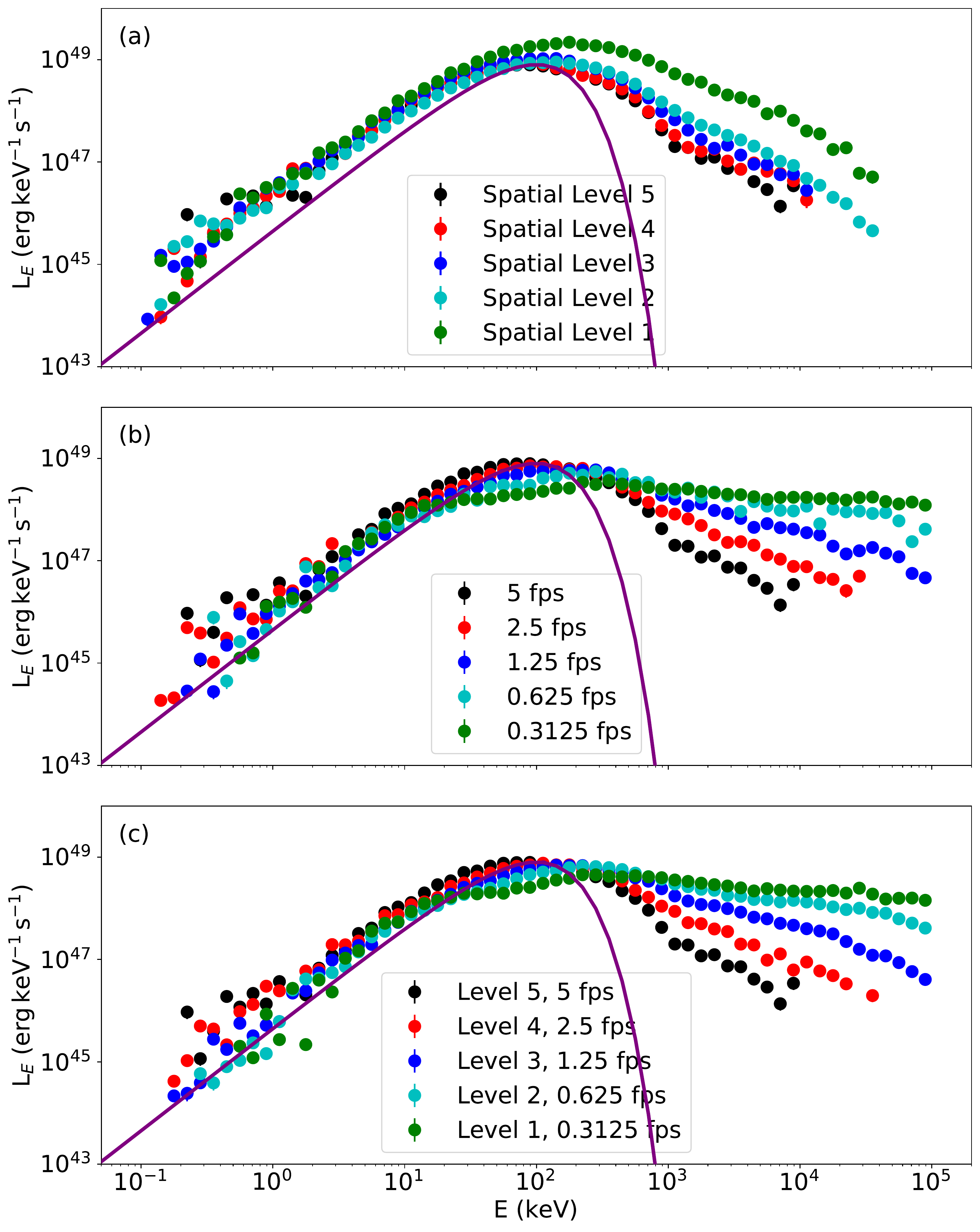}
    \caption{Spectra of a 16TI HD simulated GRB at different refinement levels. The solid purple line is a blackbody spectrum that peaks at each spectrum set's highest refinement level. Panel (a) shows the spectra of a simulated GRB using the same PLUTO 16TI simulation, but at different spatial refinement levels while maintaining the same highest temporal resolution constant. Panel (b) shows the spectra generated with the same 16TI simulation at different temporal resolutions, maintaining the highest spatial resolution constant. Panel (c) shows the spectra of a simulated GRB with matching temporal and spatial resolution levels.}
    \label{fig:spectra_science}
\end{figure}

Figure \ref{fig:spectra_science} shows spectra of a GRB simulated with a 16TI stellar progenitor model at different spatial and temporal resolutions. Figure \ref{fig:spectra_science}(a) shows the GRB spectra at the highest temporal resolution and varying spatial resolutions. Figure \ref{fig:spectra_science}(b) shows GRB spectra at the highest spatial resolution and varying temporal resolutions. Figure \ref{fig:spectra_science}(c) shows GRB spectra at matching temporal and spatial resolution levels.

Changing spatial resolutions artificially increases the spectral \replaced{high energy tail}{luminosities and peak energies} at the highest framerate, 5 fps, as is shown in Figure \ref{fig:spectra_science}(a). We see a less pronounced artificial increase in \replaced{the high energy tail}{these quantities} simulations where the framerates is held constant at a lower value then 5 fps. The introduction of a lower spatial resolution leads to a sudden change in the HD properties of each grid cell with respect to its neighbors. 
%The advantage of higher spatial resolutions is the presence of a smoother, more gradual change from cell to cell since there is a higher number of smaller grid elements at these higher resolutions. 
%A sudden change in the HD properties as the photons propagate through the HD medium and scatter with it leads to them being upscattered to higher energies.
This abrupt change in HD properties makes it hard for the photons to maintain equilibrium with the medium and leads to them being artificially upscattered. Decreased spatial resolutions lead to higher temperatures which causes an upwards shift of the spectrum, while keeping the spectral shape relatively constant. This effect is decreased as the spatial resolution is increased since the smoother HD behavior makes it easier for the photons to stay in equilibrium with the jet. The artificially increased spectral high-energy tail due to lower spatial resolutions and spurious upscatterings can be seen in Figure \ref{fig:spectra_science}(a).

%Decreasing the temporal resolution leads to the photons being artificially upscattered into higher energies, causing the spectrum to have an enhanced high energy tail. 
Lower temporal resolutions also lead to an abrupt change in the HD properties of the simulation. 
The abrupt change due to lower temporal resolutions is different in nature than the change in spatial resolutions, although the end result is the same. The lower framerate leads to photons scattering in the same HD frame for longer periods of time. Once the next frame is reached, the gradient in the HD properties is more pronounced, leading to the photons upscattering to higher energies. The artificially increased spectral high-energy tail due to lower temporal resolutions can be seen in Figure \ref{fig:spectra_science}(b). 

%\added{With decreased spatial resolutions, the higher temperatures lead to higher energies. This leads to all energy bins being artificially increased in luminosities. Since Figure \ref{fig:spectra_science}(b) shows spectra at constant spatial resolutions, we see a constant number of photons which concentrate in the higher ends of the spectrum as the framerate is decreased. This leads to a decrease in luminosity in the lower energy parts of the spectra, including near the 100 keV mark.}
Each spectrum showed in Figure \ref{fig:spectra_science}(b) is at the highest spatial resolution possible which defines the temperature of the outflow and the number of photons that are injected into the simulation. Thus, the total number of photons must be conserved in the spectrum. The dip seen in the spectra of  Figure \ref{fig:spectra_science}(b) as we lower the temporal resolution is due to the photons near the peak of the spectrum, at $\sim 100$ keV, being upscattered into the high energy tail of the spectra. At high temporal refinement levels, this dip in the spectrum at $\sim 100$ keV does not exist since the photons maintain thermal equilibrium with the jet.

Combining changes in both temporal and spatial resolutions leads to this effect being additive. There is upscattering due to large gradients in the jet's properties in both space and time however, we find that time resolution is the dominant factor in accounting for deviations from higher resolutions. This effect can be seen in Figure \ref{fig:spectra_science}(c).

\subsubsection{Lightcurves}

\begin{figure}
    \centering
    \includegraphics[width = \linewidth]{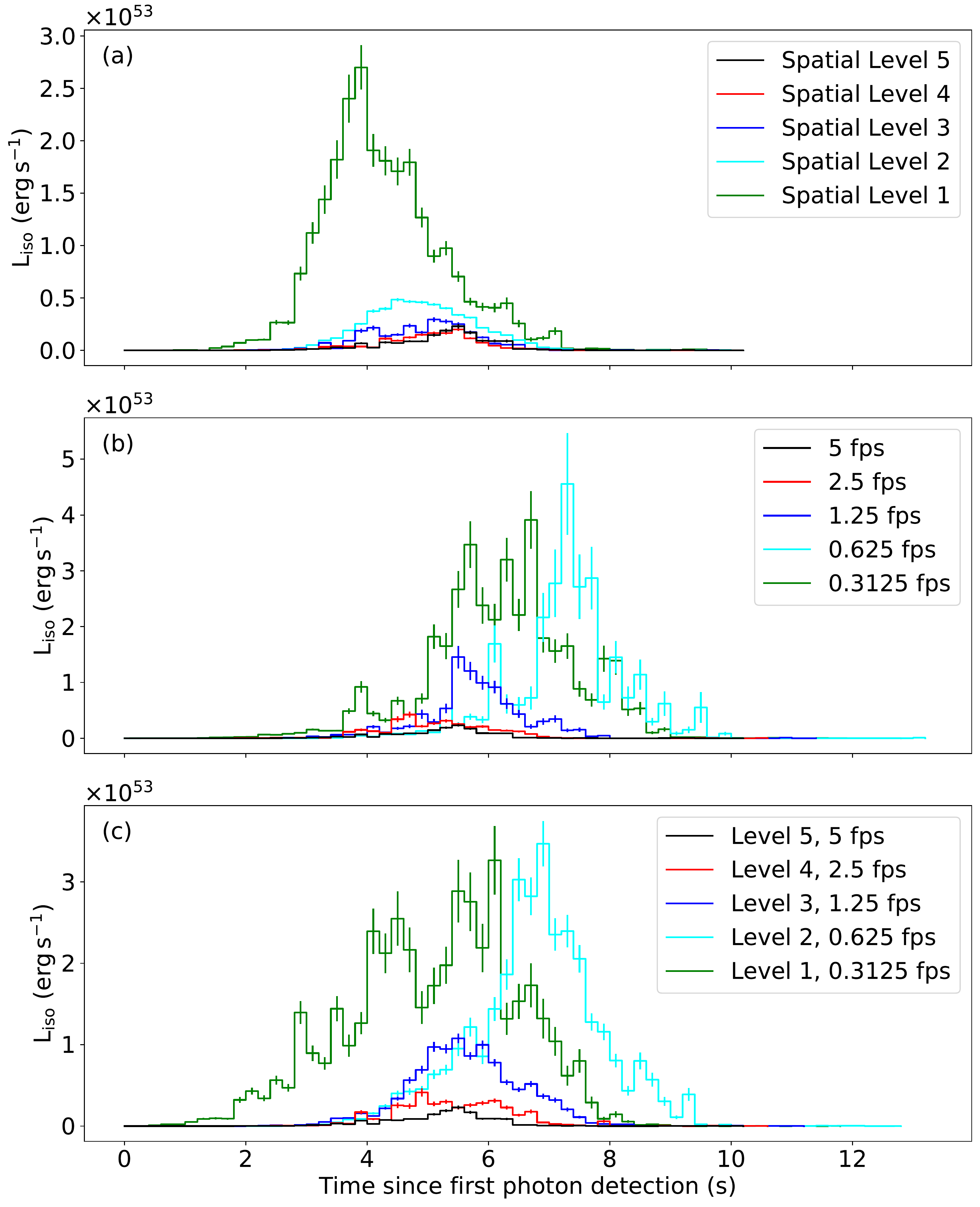}
    \caption{Lightcurves of the 16TI HD simulated GRB for different refinement levels. Only the first $\sim 10$ seconds of the simulation are shown to better visualize the qualitative effects of reducing spatial and/or temporal resolutions. Panel (a) shows the lightcurves of a simulated GRB using the same PLUTO 16TI simulation, but at different spatial refinement levels while maintaining the same highest temporal resolution constant. Panel (b) shows the lightcurves generated with the same 16TI simulation at different temporal resolutions, maintaining the highest spatial resolution constant. Panel (c) shows the lightcurves of a simulated GRB with matching temporal and spatial resolution levels. }
    \label{fig:lightcurves_science}
\end{figure}

Figure \ref{fig:lightcurves_science} shows lightcurves of a GRB simulated with a 16TI stellar progenitor model at different spatial and temporal resolutions. Only the first $\sim 10$ seconds of the lightcurve are shown to emphasize the effects of lowering temporal and/or spatial resolutions. Figure \ref{fig:lightcurves_science}(a) shows the GRB lightcurves at the highest temporal resolution and varying spatial resolutions. Figure \ref{fig:lightcurves_science}(b) shows GRB lightcurves at the highest spatial resolution and varying temporal resolutions. Figure \ref{fig:lightcurves_science}(c) shows the GRB lightcurves at matching temporal and spatial resolution levels.

Decreasing the spatial resolution for the outflow given by the 16TI simulation has a similar effect to that of decreasing the spatial resolution for a spherical outflow simulation. Because of the increase in the high energy tail of the spectrum, there is an increased luminosity that becomes more pronounced as the spatial resolution is decreased. This can be seen in Figure \ref{fig:lightcurves_science}(a). 

Changing the temporal resolution for 16TI HD simulated GRBs affects the lightcurve differently than doing so in a spherical outflow. This is due to the time dependence now present in the simulated GRB jet. Since the jet changes more drastically from frame to frame, there is an artificially enhanced high-energy tail in the spectrum, leading to a higher luminosity in the lightcurve. Not only is there a higher luminosity but the variation present in lower temporal resolutions is also present. This variation can be seen in Figure \ref{fig:lightcurves_science}(b). 

Changing both temporal and spatial resolutions leads to the combination of their individual effects. There is an increase in luminosity and in the variability of the lightcurve. The presence of the increasingly high lightcurve luminosity and variability can be seen in Figure \ref{fig:lightcurves_science}(c). 

\subsubsection{Errors in Mock Observables}

\begin{figure*}
    \centering
    \includegraphics{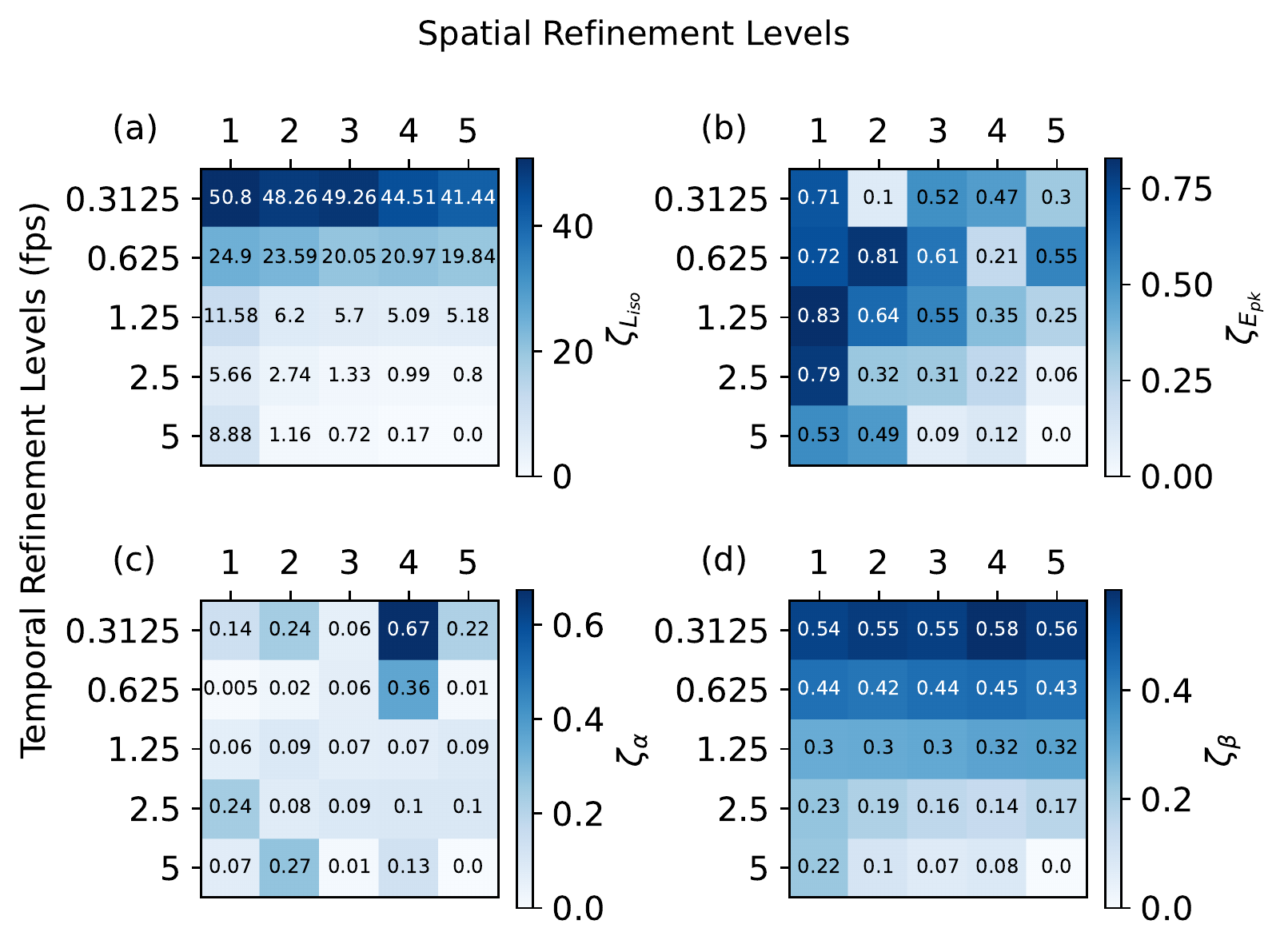}
    \caption{Matrices containing $\zeta_{\text{Prop}}$ for different EM properties of the 16TI simulated GRB. Panel (a) shows the deviation $\zeta_{L_{\text{iso}}}$ for all spatial and temporal refinement levels. Panel (b) shows the deviation $\zeta_{E_{\text{pk}}}$. Panel (c) shows the deviation $\zeta_{\alpha}$. Panel (d) shows the deviation $\zeta_{\beta}$.}
    \label{fig:zeta_all}
\end{figure*}

To better visualize the two degrees of freedom for the change in resolution, we use a color gradient matrix for all combinations of spatial and temporal resolutions. Darker colors show larger deviations from the highest level of refinement, both spatial and temporal. We find the biggest change in the lowest spatial and temporal resolutions. As we get closer to the highest resolution (refinement level 5, 5 fps), we find a smaller change in the error.

%\tyler{We want to give an overall summary of these figures, ie we see the most chnage at the lowest energy/spatial resolutions of N\% or something and as we get to the (5,5) grid point the percent errors decrease to N\%.} \red{I don't fully understand. You mean specify the maximum percent change? what does it mean as we get close? we should mention how the bottom right block has around 20\% error for example. If we move a certain number of cells up, left or diagonal from the highest resolution, we detect certain percent change.}

Luminosity shows the smallest deviation from the highest resolution at $17\%$, in the (level 4, 5 fps) entry in the resolution matrix. As we deviate from the highest resolution value in the lower right corner, we start seeing our deviation change more drastically. The largest deviation comes from the (level 1, 0.3125 fps) entry in the matrix, at a $\sim 5000\%$ change. This result is expected, since this is the lowest resolution case in both space and time.

In Figure \ref{fig:zeta_all}(a), we see $\zeta_{L_{\text{iso}}}$ for all refinement levels in the resolution matrix. The isotropic luminosity comes from integrating the spectrum with respect to energy. By looking at Figure \ref{fig:spectra_science}, we can see a clear reason why these values are changing the way that they are. The high energy tails present at lower spatial and temporal resolutions give this integral a higher value. Since there is a larger change in the spectral shape with respect to temporal resolutions, we see more similar values between spatial resolutions at constant framerates. We see an identical trend with isotropic energies $E_{\text{iso}}$. This is expected since this is an integration of the lightcurve with respect to time. 

Peak energies see deviations at around $6-55\%$, as long as we stay within the lower right $3\times 3$ block in our matrix. As we step outside of this block (lower than level 3 or 1.25 fps), we start seeing deviations of up to $83\%$.

Figure \ref{fig:zeta_all}(b) shows $\zeta_{E_\text{pk}}$ for all refinement levels in the resolution matrix for 16TI HD simulated GRBs. There is a present trend in which there is more variation present as we get further away from the lower right (level 5, 5 fps) entry. This implies that both spatial and temporal resolutions affect the resulting peak energies $E_{\text{pk}}$. The spatial resolution effect is due to the change in temperatures between levels seen in Figure \ref{fig:hdprops}. Since Equation \ref{eq:zeta_general} calculates the magnitude of the deviation and not the direction of them, these figures do not encompass this information. The peak energy values for different refinement levels are decreased as the spatial resolution is decreased.

When lowering spatial and temporal resolutions, the low energy slope, $\alpha$ varies as low as $0.05\%$ and as high as $67\%$. If we stay in the lower right $3\times 3$ block of the matrix, we see a deviation of around $1-13\%$. As we step outside of this block, (lower than level 3 or 1.25 fps), we start seeing higher deviations. The largest deviation comes from the (level 4, 0.3125 fps) entry in the matrix, at a $67\%$ change.

Figure \ref{fig:zeta_all}(c) shows $\zeta_{\alpha}$ for all refinement levels in the resolution matrix. Figure \ref{fig:spectra_science} shows some of the spectra that were evaluated here, but the trend of change in the low-energy slope $\alpha$ is not as clear by just looking at the plotted spectra. This figure shows that the lower energy slope is conserved as resolutions change, showing that lower spatial and temporal resolutions do not have a large impact on the lower energy subset of photons. Due to the scatter seen at low energies in the spectra from the low temporal resolution simulations at 0.3125 fps, as shown in Figure \ref{fig:spectra_science}(b) where there are large steps in the spectrum $\lesssim 2$ keV, the fitted $\alpha$ values are not well constrained which leads to the outliers particularly at the (level 4, 0.3125 fps) entry in the matrix. Additionally, since $\alpha$ and $E_{\text{pk}}$ are related, the low temporal resolution simulations also produce outliers in Figure \ref{fig:zeta_all}(b) see for example the (level 2, 0.3125 fps) entry of that matrix.

When lowering spatial and temporal resolutions, the high energy slope, $\beta$ is impacted more than the low energy slope $\alpha$. From Figure \ref{fig:spectra_science}, we can see that the temporal resolution affects $\beta$ more than the spatial resolution. 

Figure \ref{fig:zeta_all}(d) shows $\zeta_{\beta}$ for all refinement levels in the resolution matrix. There is an decrease in the high energy slope $\beta$ as the temporal resolutions are decreased. $\beta$ shows similar values between spatial resolutions when holding temporal resolutions constant. This effect is due to photons being upscattered to higher energies as they are shocked by drastically changing jet properties as new HD frames are loaded. The fitted Band spectrum attempts to account for the higher amount of high-energy photons by making the high energy slopes flatter.

\section{Discussions} \label{sec:discussion}

We have used MCRaT and PLUTO to quantify the effect that a simulation's spatial and temporal resolution has on post-processing radiative transfer calculations. 

We show that the lower spatial and temporal resolutions affect both the shape of the GRB's spectrum and lightcurve. The presence of high-energy spectral tails lead to an increase in isotropic luminosities as resolutions are decreased. Additionally, lower spatial resolutions lead to higher temperatures, causing photon energies to be increased, leading an increase in the normalization of the mock observed spectra.

With the lower spatial and temporal resolutions, photons propagate and scatter within a choppier HD simulation. As seen in Figure \ref{fig:hdprops}, lower spatial resolutions have higher temperatures, causing injected photons to have slightly higher energies and increased normalization for their spectrum.

As temporal resolution is decreased, the lightcurves face an increased variability, and the smoothness of this mock observed property is gone. This is due to the fact that lower framerates lead to photons not being injected as smoothly as a function of time. Higher framerates lead to a more continuous injected photon flow, while lower framerates have more spaced out photon injections. This leads to the lightcurve having more peaks and troughs compared to the higher temporal resolution simulations.

\begin{figure}
    \centering
    \includegraphics[width = \linewidth]{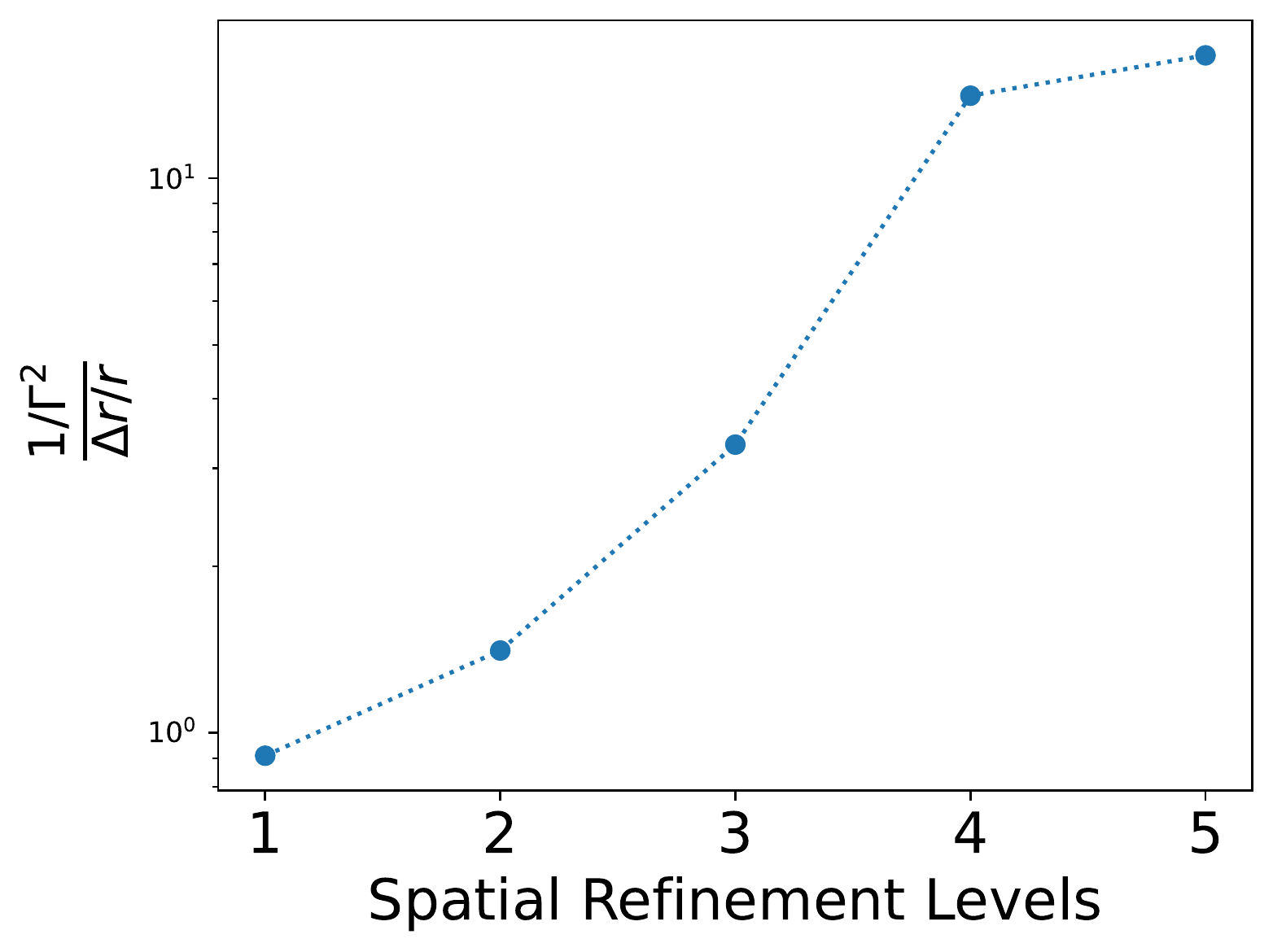}
    \caption{$\frac{1/\Gamma^2}{\Delta r / r}$ at the final frame of the 16TI HD simulation. Quantities $>>1$ mean that detected photons are sufficiently able to probe the smallest spatial scales of the jet.}
    \label{fig:theta_ratio}
\end{figure}

\begin{figure*}
    \centering
    \includegraphics[width=\linewidth]{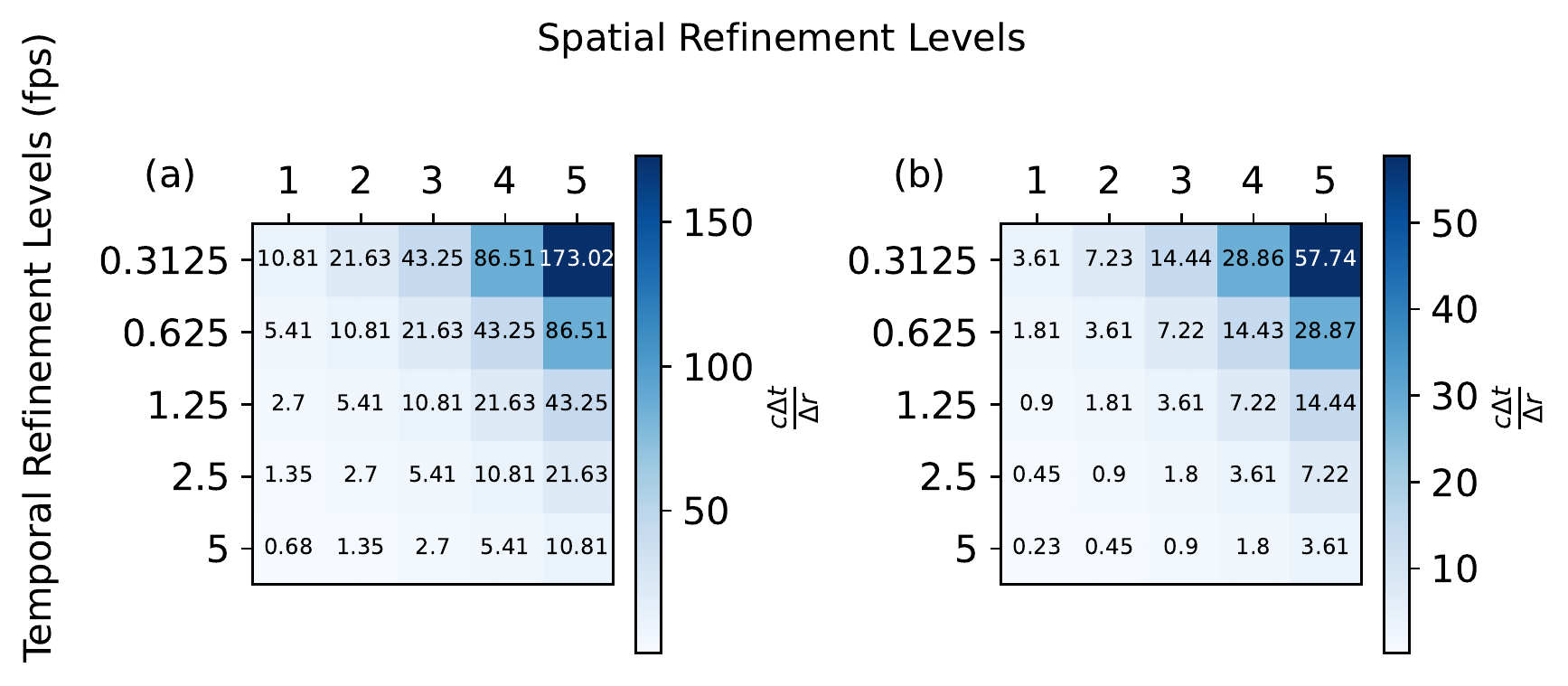}
    \caption{Lightcrossing over HD cell size ratio at beginning of the simulation. Panel(a) shows $\frac{c/fps}{\Delta r}=\frac{c\Delta t}{\Delta r}$ at the first moment of photon injection ($t_{sim} = 50$s). Panel (b) shows $\frac{c/fps}{\Delta r}=\frac{c\Delta t}{\Delta r}$ at the end of the simulation ($t_{sim} = 527.6$s).}
    \label{fig:lightcrossing}
\end{figure*}

At the end of the HD simulation, photons have ideally reached the photosphere. At this point, photons are expected to scatter into an angle of $\theta = 1/\Gamma^2$ \citep{2016lazzati}. Figure \ref{fig:hdprops}(a) shows the radial resolution at the photon position $r = 1.5\times 10^{13}$cm, and Figure \ref{fig:hdprops}(d) Shows the bulk Lorentz factor at this radius. The quantity $\Delta r / r = \sin{\theta}\approx \theta$ can be used to compare to $\theta = 1/\Gamma^2$. If the HD resolution is too low, the HD cell sizes are larger, which means that the mock observed properties of the photons are inaccurate due to the detected photons not being able to fully probe the angular scale visible to the observer. Instead photons are only probing the properties of just one large HD cell. Figure \ref{fig:theta_ratio} shows how at higher refinement levels, the ratio $\frac{1/\Gamma^2}{\Delta r / r}$ is large enough to easily probe the visible portion of the GRB jet regime since this angle encompasses multiple HD cells. At lower resolutions, we get closer but larger than unity, as seen for levels 2 and 3. Once level 1 is reached, our ratio is smaller than 1. At this resolution, the photons are not properly probing the visible angular jet region as they only probe one HD cell.

If we consider the lightcrossing defined by $c/$fps = $c\Delta t$, we get the distance traveled by a photon in between two frames at any given framerate. When comparing this distance to the size of HD cells at any given time, optimally we want the photons to scatter within multiple HD elements as they move within a given frame. For this to be true, the ratio $\frac{c\Delta t}{\Delta r}$ must be larger than unity. This would mean that the photons are able to travel through more than one HD cell during the duration of any given frame. This allows the photons to properly probe the GRB jet properties.

Figure \ref{fig:lightcrossing}(a) shows the ratio $\frac{c\Delta t}{\Delta r}$ at the beginning of the 16TI simulation. We can see that higher resolutions with lower framerates lead to photons probing multiple cells in each individual frame as they travel through the jet. The lower left corner of the matrix shows framerates and resolutions in which only one cell is probed in each frame. This phenomenon is also present and amplified at the end of the simulation, when the HD cell size is increased from the presence of higher refinement levels. This can be seen in Figure \ref{fig:lightcrossing}(b).

%We can make an analogy with polar coordinates, in which we need a larger radius in the form of lightcrossing distance, along with an angle such that there are multiple HD cells encompassed within the $1/\Gamma^2$ photon scattering angle. The angular component of the photon trajectory is affected by the spatial resolution. The resolution must be such that there are multiple HD cells encompassed within this scattering angle. For the radial component, we need a $\Delta r$ small enough such that the photon trajectory encompasses as many HD cells in each individual frame, as it travels sone distance $c \Delta t$.

We can make an analogy with polar coordinates where the photon propagates both radially, as it diffuses outward with the jet, and in polar angle, where the observer sees some amount of photons that are propagating towards them if the observer is located within the photons' local $1/\Gamma^2$ angle. As the MCRaT photon travels outward within a single simulation frame time step, we want it to interact with many cells of the GRB jet. This will allow the photon to change its properties as the jet's HD properties change radially; see for example the non-thermal spectra obtained by \cite{2018parsotanB}. Once at the photosphere, we want to receive many photons that are properly probing the angular size of the jet that the observer is able to see. This can lead to the appearance of non-thermal spectra, such as the multicolor blackbody \citep{asaf_peer2011theory}, which is typically seen in GRBs. This can only be seen if the photons are able to interact with and probe the properties of many different HD fluid elements in the polar angle direction. 

HD simulation resolution needs to be accounted for to retrieve accurate post-processing radiative transfer calculations. When deciding the HD simulation resolution for these calculations, one must look at multiple factors to make an informed decision. These factors include limitations such as storage and computational resources. Another factor that needs to be considered is what margin of error is acceptable for one's particular analysis. In order to save time, storage and computational resources, if a particular analysis allows for it, a lower resolution can be chosen, assuming the loss of accuracy in the mock EM observables is acceptable.

\begin{acknowledgments}
We thank the scientific and data editors for their helpful comments which have improved the presentation in this paper. The material is based upon work supported by NASA under award number 80GSFC21M0002. Resources supporting this work were provided by the NASA High-End Computing (HEC) Program through the NASA Advanced Supercomputing (NAS) Division at Ames Research Center.
\end{acknowledgments}

%% To help institutions obtain information on the effectiveness of their 
%% telescopes the AAS Journals has created a group of keywords for telescope 
%% facilities.
%
%% Following the acknowledgments section, use the following syntax and the
%% \facility{} or \facilities{} macros to list the keywords of facilities used 
%% in the research for the paper.  Each keyword is check against the master 
%% list during copy editing.  Individual instruments can be provided in 
%% parentheses, after the keyword, but they are not verified.

\vspace{5mm}

%% Similar to \facility{}, there is the optional \software command to allow 
%% authors a place to specify which programs were used during the creation of 
%% the manuscript. Authors should list each code and include either a
%% citation or url to the code inside ()s when available.

\software{Python \citep{python}, Astropy \citep{astropy:2013, astropy:2018, astropy:2022},  NumPy \citep{harris2020array}, MCRaT \citep{2016lazzati,2018parsotanA,2018parsotanB,parsotan2021photosphericA,parsotan2021photosphericB}, ProcessMCRaT \citep{2021processmcrat}}

%% Appendix material should be preceded with a single \appendix command.
%% There should be a \section command for each appendix. Mark appendix
%% subsections with the same markup you use in the main body of the paper.

%% Each Appendix (indicated with \section) will be lettered A, B, C, etc.
%% The equation counter will reset when it encounters the \appendix
%% command and will number appendix equations (A1), (A2), etc. The
%% Figure and Table counter will not reset.

%\appendix

%\section{Appendix information}

%\section{Gold Open Access}

%\section{Author publication charges}

%\section{Rotating tables}

\bibliography{sample631}{}

\begin{thebibliography}{}
\expandafter\ifx\csname natexlab\endcsname\relax\def\natexlab#1{#1}\fi
\providecommand{\url}[1]{\href{#1}{#1}}
\providecommand{\dodoi}[1]{doi:~\href{http://doi.org/#1}{\nolinkurl{#1}}}
\providecommand{\doeprint}[1]{\href{http://ascl.net/#1}{\nolinkurl{http://ascl.net/#1}}}
\providecommand{\doarXiv}[1]{\href{https://arxiv.org/abs/#1}{\nolinkurl{https://arxiv.org/abs/#1}}}

\bibitem[{{Amati} {et~al.}(2002){Amati}, {Frontera}, {Tavani}, {in't Zand},
  {Antonelli}, {Costa}, {Feroci}, {Guidorzi}, {Heise}, {Masetti}, {Montanari},
  {Nicastro}, {Palazzi}, {Pian}, {Piro}, \& {Soffitta}}]{2002amati}
{Amati}, L., {Frontera}, F., {Tavani}, M., {et~al.} 2002, \aap, 390, 81,
  \dodoi{10.1051/0004-6361:20020722}

\bibitem[{{Astropy Collaboration} {et~al.}(2013){Astropy Collaboration},
  {Robitaille}, {Tollerud}, {Greenfield}, {Droettboom}, {Bray}, {Aldcroft},
  {Davis}, {Ginsburg}, {Price-Whelan}, {Kerzendorf}, {Conley}, {Crighton},
  {Barbary}, {Muna}, {Ferguson}, {Grollier}, {Parikh}, {Nair}, {Unther},
  {Deil}, {Woillez}, {Conseil}, {Kramer}, {Turner}, {Singer}, {Fox}, {Weaver},
  {Zabalza}, {Edwards}, {Azalee Bostroem}, {Burke}, {Casey}, {Crawford},
  {Dencheva}, {Ely}, {Jenness}, {Labrie}, {Lim}, {Pierfederici}, {Pontzen},
  {Ptak}, {Refsdal}, {Servillat}, \& {Streicher}}]{astropy:2013}
{Astropy Collaboration}, {Robitaille}, T.~P., {Tollerud}, E.~J., {et~al.} 2013,
  \aap, 558, A33, \dodoi{10.1051/0004-6361/201322068}

\bibitem[{{Astropy Collaboration} {et~al.}(2018){Astropy Collaboration},
  {Price-Whelan}, {Sip{\H{o}}cz}, {G{\"u}nther}, {Lim}, {Crawford}, {Conseil},
  {Shupe}, {Craig}, {Dencheva}, {Ginsburg}, {Vand erPlas}, {Bradley},
  {P{\'e}rez-Su{\'a}rez}, {de Val-Borro}, {Aldcroft}, {Cruz}, {Robitaille},
  {Tollerud}, {Ardelean}, {Babej}, {Bach}, {Bachetti}, {Bakanov}, {Bamford},
  {Barentsen}, {Barmby}, {Baumbach}, {Berry}, {Biscani}, {Boquien}, {Bostroem},
  {Bouma}, {Brammer}, {Bray}, {Breytenbach}, {Buddelmeijer}, {Burke},
  {Calderone}, {Cano Rodr{\'\i}guez}, {Cara}, {Cardoso}, {Cheedella}, {Copin},
  {Corrales}, {Crichton}, {D'Avella}, {Deil}, {Depagne}, {Dietrich}, {Donath},
  {Droettboom}, {Earl}, {Erben}, {Fabbro}, {Ferreira}, {Finethy}, {Fox},
  {Garrison}, {Gibbons}, {Goldstein}, {Gommers}, {Greco}, {Greenfield},
  {Groener}, {Grollier}, {Hagen}, {Hirst}, {Homeier}, {Horton}, {Hosseinzadeh},
  {Hu}, {Hunkeler}, {Ivezi{\'c}}, {Jain}, {Jenness}, {Kanarek}, {Kendrew},
  {Kern}, {Kerzendorf}, {Khvalko}, {King}, {Kirkby}, {Kulkarni}, {Kumar},
  {Lee}, {Lenz}, {Littlefair}, {Ma}, {Macleod}, {Mastropietro}, {McCully},
  {Montagnac}, {Morris}, {Mueller}, {Mumford}, {Muna}, {Murphy}, {Nelson},
  {Nguyen}, {Ninan}, {N{\"o}the}, {Ogaz}, {Oh}, {Parejko}, {Parley}, {Pascual},
  {Patil}, {Patil}, {Plunkett}, {Prochaska}, {Rastogi}, {Reddy Janga},
  {Sabater}, {Sakurikar}, {Seifert}, {Sherbert}, {Sherwood-Taylor}, {Shih},
  {Sick}, {Silbiger}, {Singanamalla}, {Singer}, {Sladen}, {Sooley},
  {Sornarajah}, {Streicher}, {Teuben}, {Thomas}, {Tremblay}, {Turner},
  {Terr{\'o}n}, {van Kerkwijk}, {de la Vega}, {Watkins}, {Weaver}, {Whitmore},
  {Woillez}, {Zabalza}, \& {Astropy Contributors}}]{astropy:2018}
{Astropy Collaboration}, {Price-Whelan}, A.~M., {Sip{\H{o}}cz}, B.~M., {et~al.}
  2018, \aj, 156, 123, \dodoi{10.3847/1538-3881/aabc4f}

\bibitem[{{Astropy Collaboration} {et~al.}(2022){Astropy Collaboration},
  {Price-Whelan}, {Lim}, {Earl}, {Starkman}, {Bradley}, {Shupe}, {Patil},
  {Corrales}, {Brasseur}, {N{"o}the}, {Donath}, {Tollerud}, {Morris},
  {Ginsburg}, {Vaher}, {Weaver}, {Tocknell}, {Jamieson}, {van Kerkwijk},
  {Robitaille}, {Merry}, {Bachetti}, {G{"u}nther}, {Aldcroft},
  {Alvarado-Montes}, {Archibald}, {B{'o}di}, {Bapat}, {Barentsen}, {Baz{'a}n},
  {Biswas}, {Boquien}, {Burke}, {Cara}, {Cara}, {Conroy}, {Conseil}, {Craig},
  {Cross}, {Cruz}, {D'Eugenio}, {Dencheva}, {Devillepoix}, {Dietrich},
  {Eigenbrot}, {Erben}, {Ferreira}, {Foreman-Mackey}, {Fox}, {Freij}, {Garg},
  {Geda}, {Glattly}, {Gondhalekar}, {Gordon}, {Grant}, {Greenfield}, {Groener},
  {Guest}, {Gurovich}, {Handberg}, {Hart}, {Hatfield-Dodds}, {Homeier},
  {Hosseinzadeh}, {Jenness}, {Jones}, {Joseph}, {Kalmbach}, {Karamehmetoglu},
  {Ka{l}uszy{'n}ski}, {Kelley}, {Kern}, {Kerzendorf}, {Koch}, {Kulumani},
  {Lee}, {Ly}, {Ma}, {MacBride}, {Maljaars}, {Muna}, {Murphy}, {Norman},
  {O'Steen}, {Oman}, {Pacifici}, {Pascual}, {Pascual-Granado}, {Patil},
  {Perren}, {Pickering}, {Rastogi}, {Roulston}, {Ryan}, {Rykoff}, {Sabater},
  {Sakurikar}, {Salgado}, {Sanghi}, {Saunders}, {Savchenko}, {Schwardt},
  {Seifert-Eckert}, {Shih}, {Jain}, {Shukla}, {Sick}, {Simpson},
  {Singanamalla}, {Singer}, {Singhal}, {Sinha}, {Sip{H{o}}cz}, {Spitler},
  {Stansby}, {Streicher}, {{{S}}umak}, {Swinbank}, {Taranu}, {Tewary},
  {Tremblay}, {Val-Borro}, {Van Kooten}, {Vasovi{'c}}, {Verma}, {de Miranda
  Cardoso}, {Williams}, {Wilson}, {Winkel}, {Wood-Vasey}, {Xue}, {Yoachim},
  {Zhang}, {Zonca}, \& {Astropy Project Contributors}}]{astropy:2022}
{Astropy Collaboration}, {Price-Whelan}, A.~M., {Lim}, P.~L., {et~al.} 2022,
  apj, 935, 167, \dodoi{10.3847/1538-4357/ac7c74}

\bibitem[{{Band} {et~al.}(1993){Band}, {Matteson}, {Ford}, {Schaefer},
  {Palmer}, {Teegarden}, {Cline}, {Briggs}, {Paciesas}, {Pendleton}, {Fishman},
  {Kouveliotou}, {Meegan}, {Wilson}, \& {Lestrade}}]{1993Band}
{Band}, D., {Matteson}, J., {Ford}, L., {et~al.} 1993, \apj, 413, 281,
  \dodoi{10.1086/172995}

\bibitem[{{Beloborodov}(2010)}]{2010beleborodov}
{Beloborodov}, A.~M. 2010, \mnras, 407, 1033,
  \dodoi{10.1111/j.1365-2966.2010.16770.x}

\bibitem[{{Chhotray} \& {Lazzati}(2015)}]{2015chhotray-lazzati}
{Chhotray}, A., \& {Lazzati}, D. 2015, \apj, 802, 132,
  \dodoi{10.1088/0004-637X/802/2/132}

\bibitem[{{Daigne} {et~al.}(2011){Daigne}, {Bo{\v{s}}njak}, \&
  {Dubus}}]{2011daigne}
{Daigne}, F., {Bo{\v{s}}njak}, {\v{Z}}., \& {Dubus}, G. 2011, \aap, 526, A110,
  \dodoi{10.1051/0004-6361/201015457}

\bibitem[{Harris {et~al.}(2020)Harris, Millman, van~der Walt, Gommers,
  Virtanen, Cournapeau, Wieser, Taylor, Berg, Smith, Kern, Picus, Hoyer, van
  Kerkwijk, Brett, Haldane, del R{\'{i}}o, Wiebe, Peterson,
  G{\'{e}}rard-Marchant, Sheppard, Reddy, Weckesser, Abbasi, Gohlke, \&
  Oliphant}]{harris2020array}
Harris, C.~R., Millman, K.~J., van~der Walt, S.~J., {et~al.} 2020, Nature, 585,
  357, \dodoi{10.1038/s41586-020-2649-2}

\bibitem[{Ito {et~al.}(2018)Ito, Levinson, Stern, \& Nagataki}]{ito_rms}
Ito, H., Levinson, A., Stern, B.~E., \& Nagataki, S. 2018, Monthly Notices of
  the Royal Astronomical Society, 474, 2828

\bibitem[{{Ito} {et~al.}(2015){Ito}, {Matsumoto}, {Nagataki}, {Warren}, \&
  {Barkov}}]{2015ito}
{Ito}, H., {Matsumoto}, J., {Nagataki}, S., {Warren}, D.~C., \& {Barkov}, M.~V.
  2015, \apjl, 814, L29, \dodoi{10.1088/2041-8205/814/2/L29}

\bibitem[{{Ito} {et~al.}(2019){Ito}, {Matsumoto}, {Nagataki}, {Warren},
  {Barkov}, \& {Yonetoku}}]{2019ito}
{Ito}, H., {Matsumoto}, J., {Nagataki}, S., {et~al.} 2019, Nature
  Communications, 10, 1504, \dodoi{10.1038/s41467-019-09281-z}

\bibitem[{{Ito} {et~al.}(2014){Ito}, {Nagataki}, {Matsumoto}, {Lee}, {Tolstov},
  {Mao}, {Dainotti}, \& {Mizuta}}]{2014ito}
{Ito}, H., {Nagataki}, S., {Matsumoto}, J., {et~al.} 2014, \apj, 789, 159,
  \dodoi{10.1088/0004-637X/789/2/159}

\bibitem[{{Ito} {et~al.}(2013){Ito}, {Nagataki}, {Ono}, {Lee}, {Mao}, {Yamada},
  {Pe'er}, {Mizuta}, \& {Harikae}}]{2013ito}
{Ito}, H., {Nagataki}, S., {Ono}, M., {et~al.} 2013, \apj, 777, 62,
  \dodoi{10.1088/0004-637X/777/1/62}

\bibitem[{{Klebesadel} {et~al.}(1973){Klebesadel}, {Strong}, \&
  {Olson}}]{1973klebesadel}
{Klebesadel}, R.~W., {Strong}, I.~B., \& {Olson}, R.~A. 1973, \apjl, 182, L85,
  \dodoi{10.1086/181225}

\bibitem[{{Lazzati}(2016)}]{2016lazzati}
{Lazzati}, D. 2016, \apj, 829, 76, \dodoi{10.3847/0004-637X/829/2/76}

\bibitem[{{Lazzati} {et~al.}(2009){Lazzati}, {Morsony}, \&
  {Begelman}}]{2009lazzati}
{Lazzati}, D., {Morsony}, B.~J., \& {Begelman}, M.~C. 2009, \apjl, 700, L47,
  \dodoi{10.1088/0004-637X/700/1/L47}

\bibitem[{{Lazzati} {et~al.}(2013){Lazzati}, {Morsony}, {Margutti}, \&
  {Begelman}}]{2013lazzatiA}
{Lazzati}, D., {Morsony}, B.~J., {Margutti}, R., \& {Begelman}, M.~C. 2013,
  \apj, 765, 103, \dodoi{10.1088/0004-637X/765/2/103}

\bibitem[{Lazzati {et~al.}(2013)Lazzati, Morsony, Margutti, \&
  Begelman}]{lazzati2013photospheric}
Lazzati, D., Morsony, B.~J., Margutti, R., \& Begelman, M.~C. 2013, The
  Astrophysical Journal, 765, 103

\bibitem[{{L{\'o}pez-C{\'a}mara} {et~al.}(2014){L{\'o}pez-C{\'a}mara},
  {Morsony}, \& {Lazzati}}]{2014lopez-camara}
{L{\'o}pez-C{\'a}mara}, D., {Morsony}, B.~J., \& {Lazzati}, D. 2014, \mnras,
  442, 2202, \dodoi{10.1093/mnras/stu1016}

\bibitem[{{Mignone} {et~al.}(2007){Mignone}, {Bodo}, {Massaglia}, {Matsakos},
  {Tesileanu}, {Zanni}, \& {Ferrari}}]{2007mignone}
{Mignone}, A., {Bodo}, G., {Massaglia}, S., {et~al.} 2007, \apjs, 170, 228,
  \dodoi{10.1086/513316}

\bibitem[{{Mignone} {et~al.}(2012){Mignone}, {Zanni}, {Tzeferacos}, {van
  Straalen}, {Colella}, \& {Bodo}}]{2012mignone}
{Mignone}, A., {Zanni}, C., {Tzeferacos}, P., {et~al.} 2012, \apjs, 198, 7,
  \dodoi{10.1088/0067-0049/198/1/7}

\bibitem[{{Parsotan}(2021)}]{2021processmcrat}
{Parsotan}, T. 2021, {parsotat/ProcessMCRaT: Cyclo-Synchrotron Release},
  v1.0.1, Zenodo,  Zenodo, \dodoi{10.5281/zenodo.4918108}

\bibitem[{{Parsotan} {et~al.}(2021){Parsotan}, {Cochrane}, {Hayward},
  {Angl{\'e}s-Alc{\'a}zar}, {Feldmann}, {Faucher-Gigu{\`e}re}, {Wellons}, \&
  {Hopkins}}]{parsotan_galaxy_rt}
{Parsotan}, T., {Cochrane}, R.~K., {Hayward}, C.~C., {et~al.} 2021, \mnras,
  501, 1591, \dodoi{10.1093/mnras/staa3765}

\bibitem[{{Parsotan} \& {Lazzati}(2018)}]{2018parsotanA}
{Parsotan}, T., \& {Lazzati}, D. 2018, \apj, 853, 8,
  \dodoi{10.3847/1538-4357/aaa087}

\bibitem[{Parsotan \& Lazzati(2021)}]{parsotan2021photosphericA}
Parsotan, T., \& Lazzati, D. 2021, The Astrophysical Journal, 922, 257

\bibitem[{{Parsotan} \& {Lazzati}(2022)}]{parsotan2021photosphericB}
{Parsotan}, T., \& {Lazzati}, D. 2022, \apj, 926, 104,
  \dodoi{10.3847/1538-4357/ac4093}

\bibitem[{{Parsotan} {et~al.}(2018){Parsotan}, {L{\'o}pez-C{\'a}mara}, \&
  {Lazzati}}]{2018parsotanB}
{Parsotan}, T., {L{\'o}pez-C{\'a}mara}, D., \& {Lazzati}, D. 2018, \apj, 869,
  103, \dodoi{10.3847/1538-4357/aaeed1}

\bibitem[{{Pe'er}(2008)}]{2008peer}
{Pe'er}, A. 2008, \apj, 682, 463, \dodoi{10.1086/588136}

\bibitem[{{Pe'er} \& {Ryde}(2011)}]{2011peer-ryde}
{Pe'er}, A., \& {Ryde}, F. 2011, \apj, 732, 49,
  \dodoi{10.1088/0004-637X/732/1/49}

\bibitem[{Pe'er \& Ryde(2011)}]{asaf_peer2011theory}
Pe'er, A., \& Ryde, F. 2011, The Astrophysical Journal, 732, 49

\bibitem[{{Python Core Team}(2019)}]{python}
{Python Core Team}. 2019, {Python: A dynamic, open source programming
  language}, {Python Software Foundation}.
\newblock \url{https://www.python.org/}

\bibitem[{{Rees} \& {Meszaros}(1994)}]{1994meszaros}
{Rees}, M.~J., \& {Meszaros}, P. 1994, \apjl, 430, L93, \dodoi{10.1086/187446}

\bibitem[{{Rees} \& {M{\'e}sz{\'a}ros}(2005)}]{2005meszaros}
{Rees}, M.~J., \& {M{\'e}sz{\'a}ros}, P. 2005, \apj, 628, 847,
  \dodoi{10.1086/430818}

\bibitem[{Stamatellos \& Whitworth(2005)}]{sph_rt}
Stamatellos, D., \& Whitworth, A.~P. 2005, Astronomy \& Astrophysics, 439, 153

\bibitem[{{Vurm} \& {Beloborodov}(2016)}]{2016vurm}
{Vurm}, I., \& {Beloborodov}, A.~M. 2016, \apj, 831, 175,
  \dodoi{10.3847/0004-637X/831/2/175}

\bibitem[{Woosley \& Heger(2006)}]{woosley2006progenitor}
Woosley, S., \& Heger, A. 2006, The Astrophysical Journal, 637, 914

\bibitem[{{Yonetoku} {et~al.}(2004){Yonetoku}, {Murakami}, {Nakamura},
  {Yamazaki}, {Inoue}, \& {Ioka}}]{2004yonetoku}
{Yonetoku}, D., {Murakami}, T., {Nakamura}, T., {et~al.} 2004, \apj, 609, 935,
  \dodoi{10.1086/421285}

\bibitem[{{Zhang} \& {Yan}(2011)}]{2011zhang-yan}
{Zhang}, B., \& {Yan}, H. 2011, \apj, 726, 90,
  \dodoi{10.1088/0004-637X/726/2/90}

\end{thebibliography}
\bibliographystyle{aasjournal}

%% This command is needed to show the entire author+affiliation list when
%% the collaboration and author truncation commands are used.  It has to
%% go at the end of the manuscript.
%\allauthors

%% Include this line if you are using the \added, \replaced, \deleted
%% commands to see a summary list of all changes at the end of the article.
\listofchanges

\end{document}